# Extinction of incident hydrogen/air detonation in fine water sprays

Yong Xu, Majie Zhao, Huangwei Zhang*

*Department of Mechanical Engineering, National University of Singapore, 9 Engineering Drive 1, Singapore, 117576, Republic of Singapore*

## Abstract

Two-dimensional numerical simulations with Eulerian-Lagrangian method are conducted to study propagation and extinction of stoichiometric hydrogen/air detonations in fine water sprays. Parameterized by water mass loading and initial droplet size, a detonation extinction map is developed. Detonation extinction would occur with larger mass loading and/or smaller droplet size. General features of gas phase and water droplets and local detonation frontal structures are well captured. Numerical soot foils are used to characterize the influence of mass loading and droplet size on the detonation wave. The results also show that the detonation cell size increases with increased mass loading or decreased droplet size. Analysis on unsteady detonation extinction process is performed with the evolutions of detonation frontal structure, spatial distribution of thermochemical variables and interphase transfer rates (mass, energy, and momentum). Moreover, the chemical explosive mode analysis reveals that for stable detonation, thermal runaway dominates behind the Mach stem, while chemical propensities of auto-ignition and thermal runaway appear alternately behind the incident wave. When the induction zone length increases as the reaction front (RF) and shock front (SF) are decoupled, localized burned pockets surrounded by the autoignition chemical explosive mode can be observed. In addition, the interactions between detonation wave and water droplets demonstrate that the energy and momentum transfer have more direct interaction with SF and RF than the mass transfer. The interphase transfer rates increase with the water mass loading. Under the same mass loading, the smaller the droplet size, the larger the interphase transfer rates. However, the size of fine water droplets has a limited influence on the interphase momentum exchange. Moreover, high energy and mass transfer rates are observed at the onset of detonation extinction, and they gradually decrease when the reaction and detonation fronts are decoupled.

* Corresponding author. Tel.: +65 6516 2557; Fax: +65 6779 1459.
*E-mail address*: huangwei.zhang@nus.edu.sg.

## 1. Introduction

There are increased interests in exploring effective approaches to mitigate detonation of flammable gas, related to prevention from explosion hazards and industrial safety assessment [1,2]. Water is an ideal detonation mitigant due to numerous advantages [3,4]. Specifically, it can absorb considerable heat from gas phase due to large heat capacity and latent heat of evaporation [5]. Also, it is readily available with low cost and lots of flexibilities. Meanwhile, use of water would not bring environmental pollution. Moreover, as a solution, it is possible to include proper additives, e.g., alkali salts (KCl and NaCl) [6], for better explosion inhibition. There are various forms of water utilized for detonation or explosion mitigation [5,7,8], e.g., solid jet, diffuse jet and water mists. The last one is most promising since sprayed water droplets have large specific surface area and low terminal velocity, which allow them to continuously circulate in the explosion area in a manner of a total flooding gas. It therefore can effectively weaken the blast, inhibit chemical reaction, and reduce gas temperature. Although it has been widely used in various areas, e.g., process and energy industries, nuclear power plants, and even space applications, however, the mechanisms behind detonation / explosion inhibition with water mists are still not well understood.

There have been a series of studies about propagation of shock / blast waves in water sprays. For instance, Jourdan et al. [9] use water aerosol shock tube experiments to study shock attenuation in a cloud of water droplets. They characterize shock attenuation with shock tube (i.e., cross-sectional area) and droplet properties (e.g., total water volume, droplet size, loading rate and droplet specific surface area). With the similar experimental conditions, Chauvin et al. [10] find the peculiar pressure evolution after the transmitted shock wave in two-phase mixture and they also measure the overpressures under different water spray conditions. Moreover, Adiga et al. [11] unveil the physical picture of fine water droplet breakup in detonation process and quantify the droplet fragmentation with breakup energy. Eulerian–Lagrangian simulations by Ananth et al. [12] are performed to examine the effects of mono-dispersed fine water mist on a confined blast. It is found that the latent heat absorption is dominant for blast mitigation, followed by convective heat transfer and momentum exchange. Furthermore, Schwer and Kailasanath [13] simulate unconfined explosions in water sprays, and find that the water mists can dampen the shock through vaporization and momentum extraction. Different from the observations by Ananth et al. [12], they claim that the momentum extraction plays a more important role in weakening the blast.

In the abovementioned studies, the effects of water mists on chemical reactions are not discussed, since they use air as the carrier gas (e.g., [9,10]), specify the chemically equilibrium gas from an explosion (e.g., [12]), or there is no direct interaction between water droplets and post-shock reaction zone (e.g. [13]). It is well known that detonation is a complex of coupled shock and reaction fronts, and therefore additional complexities may arise in droplet−detonation interactions. Thomas et al. [14] experimentally study detonations of hydrogen, ethane and acetylene with water sprays in a vertical tube. They attribute detonation failure to high heat loss due to water droplets compared to the combustion heat release. They also find that the water droplet size and loading densities are crucial to quenching a detonation. It is observed by Niedzielska et al. [15] that small (215 μm in their experiments) droplets with fast evaporation rate has strong influence on detonation quenching. Moreover, from detonation tube experiments, Jarsalé et al. [16] observe that presence of water spray drastically alters the detonation cell size, but the ratio of the hydrodynamic thickness to the cell size remain constant, regardless of water droplet addition. The effects of water mists on Deflagration-to-Detonation Transition (DDT) are also demonstrated with reduced overpressure and delayed timing of detonation development [8].

Besides the foregoing experimental work [8,14–16], recent computational efforts provide us more insights on detonation in water sprays. For example, Song and Zhang simulate the methane detonation and find that the inhibition effects of water sprays are mainly reflected in reduction of flame temperature [7]. Watanabe et al. observe that the dispersed water droplets significantly alter the detonation flow field, and water droplet evaporation mainly occurs around 10 mm behind the leading shock [17]. Furthermore, the cellular patterns of dilute water spray detonation are more regular than those of the droplet-free detonations [18]. The interactions between detonation wave and water droplets change the hydrodynamic thickness. Exchange of mass, momentum, and energy between two phases occurs within the hydrodynamic thickness, which lowers the detonation velocity and fluctuations downstream of the leading shock front. Their results also show that droplet breakup mainly occurs near the shock front [19], and the average diameter of the disintegrated water droplets is independent on the initial propagation velocity of the shock front. However, in these numerical studies, detonation extinction due to water sprays are not investigated.

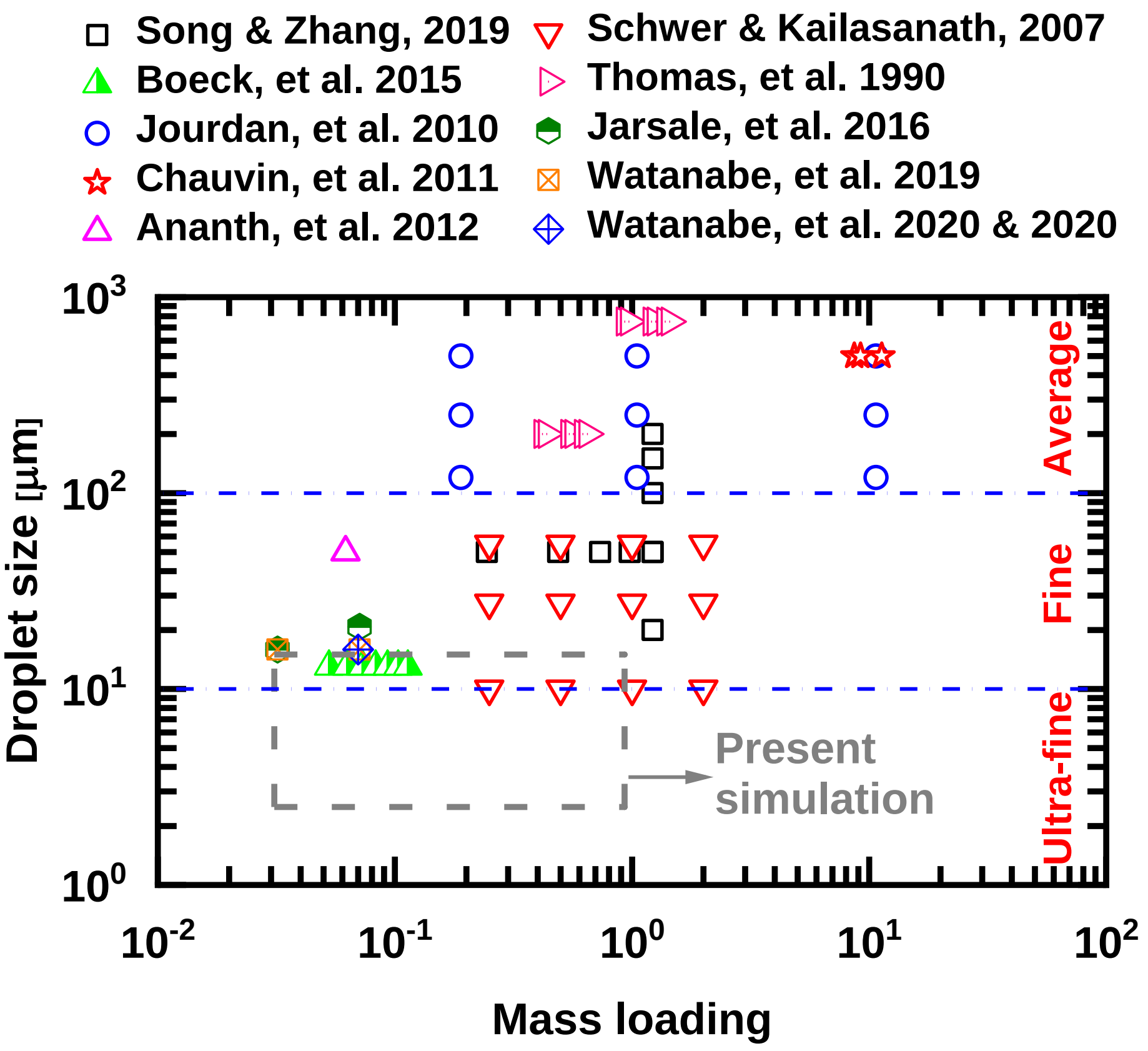

Figure 1 Studies of detonation and shock in water sprays. Droplet diameter spectrum (dash-dotted lines) follows Grant et al. [5]. Average: droplet diameter range most relevant for firefighting.

The droplet size and mass loading considered in the above studies are summarized in Fig. 1. One can see that most of the droplet diameters are above 20 μm, corresponding to the mass loading of 0.03−13.3. It is well known that fine or ultra-fine water droplets with diameter less than 20 μm have outstanding performance in fire suppression, due to fast evaporation rate and high specific surface area [3,4]. Nevertheless, their effectiveness and the critical spray properties for detonation extinction and how the sprayed droplets interact with the detonation have not been reported yet. In this work, propagation and extinction of incident hydrogen/air detonation in fine or ultra-fine water droplets will be computationally studied. As marked in Fig. 1, the droplet diameters considered in the present simulations range from 2.5 to 15 μm, whilst the mass loading is 0.031—0.93. The Eulerian−Lagrangian approach with two-way gas−liquid coupling is used to model the compressible, multispecies, and two-phase reacting flows. The Chemical Explosive Mode Analysis (CEMA) [20,21] is applied to extract the detailed information about the chemical reaction in gaseous detonations. Emphasis is laid on the interactions between gas phase and droplet phase, as well as the chemical structure evolutions when the hydrogen/air detonation extinction process. The rest of the manuscript is structured as below. The governing equation and numerical implementations will be presented in Section 2, whilst the physical model will be detailed in Section 3. The results will be presented in Section 4, followed by the discussion in Section 5. The main findings are summarized in Section 6.

## 2. Governing equation and computational method

### *2.1 Gas phase*

The governing equations of mass, momentum, energy, and species mass fraction are solved with the ideal gas equation of state. They respectively read

$$\frac{\partial \rho}{\partial t} + \nabla \cdot [\rho \mathbf{u}] = S_{mass}, \tag{1}$$

$$\frac{\partial (\rho \mathbf{u})}{\partial t} + \nabla \cdot [\mathbf{u}(\rho \mathbf{u})] + \nabla p + \nabla \cdot \mathbf{T} = \mathbf{S}_{mom}, \tag{2}$$

$$\frac{\partial (\rho \boldsymbol{E})}{\partial t} + \nabla \cdot [\mathbf{u}(\rho \boldsymbol{E} + \boldsymbol{p})] + \nabla \cdot [\mathbf{T} \cdot \mathbf{u}] + \nabla \cdot \mathbf{j} = \dot{\omega}_T + S_{energy}, \tag{3}$$

$$\frac{\partial (\rho Y_m)}{\partial t} + \nabla \cdot [\mathbf{u}(\rho Y_m)] + \nabla \cdot \mathbf{s_m} = \dot{\omega}_m + S_{species,m}, (m = 1, \dots M-1), \tag{4}$$

$$p = \rho R T. \tag{5}$$

In above equations, $t$ is time and $\nabla \cdot (\cdot)$ is the divergence operator. $\rho$ is the gas density, $\mathbf{u}$ is the velocity vector, $T$ is the gas temperature, and $p$ is the pressure, which is updated from the equation of state, i.e. Eq. (5). $Y_m$ is the mass fraction of $m$-th species, and $M$ is the total species number. Only ($M-1$) equations are solved in Eq. (4) and the mass fraction of the inert species (e.g., nitrogen) is recovered from $\sum_{m=1}^{M} Y_m = 1$. $\boldsymbol{E} \equiv e + |\mathbf{u}|^2/2$ is the total energy, and $\boldsymbol{e}$ is the specific internal energy. $R$ in Eq. (5) is the specific gas constant and is calculated from $R = R_u \sum_{m=1}^{M} Y_m W_m^{-1}$. $W_m$ is the molar weight of $m$-th species and $R_u = 8.314$ J/(mol·K) is the universal gas constant. The source terms in Eqs. (1)−(4), i.e. $S_{mass}$, $\mathbf{S}_{mom}$, $S_{energy}$ and $S_{species,m}$, denote the exchanges of mass, momentum, energy and species between gas and liquid phases. Their corresponding expressions are given in Eqs. (26)−(29), respectively.

The viscous stress tensor $\mathbf{T}$ in Eq. (2) is modelled by

$$\mathbf{T} = -2\mu \mathrm{dev}(\mathbf{D}) . \quad (6)$$

Here $\mu$ is the dynamic viscosity and follows the Sutherland's law [22]. Moreover, $\mathrm{dev}(\mathbf{D}) = \mathbf{D} - \mathrm{tr}(\mathbf{D})\mathbf{I}/3$ is the deviatoric component of the deformation gradient tensor $\mathbf{D} \equiv [\nabla\mathbf{u} + (\nabla\mathbf{u})^T]/2$. $\mathbf{I}$ denotes the unit tensor.

In addition, $\mathbf{j}$ in Eq. (3) is the diffusive heat flux and can be modelled with Fourier's law, i.e.

$$\mathbf{j} = -k\nabla T. \quad (7)$$

Thermal conductivity $k$ is calculated using the Eucken approximation [23], i.e., $k = \mu C_v(1.32 + 1.37R/C_v)$, where $C_v$ is the heat capacity at constant volume and derived from $C_v = C_p - R$. Here $C_p = \sum_{m=1}^{M} Y_m C_{p,m}$ is the heat capacity at constant pressure, and $C_{p,m}$ is estimated from JANAF polynomials [24].

In Eq. (4), $\mathbf{s_m} = -D_m \nabla(\rho Y_m)$ is the species mass flux. $D_m = \alpha/Le_m$ is the mass diffusivity. With the unity Lewis number assumption (i.e., $Le_m = 1$), $D_m$ can be calculated through $D_m = k/\rho C_p$. Moreover, $\dot{\omega}_m$ is the production or consumption rate of $m$-th species by all $N$ reactions

$$\dot{\omega}_m = W_m \sum_{j=1}^{N} \omega_{m,j}^o . \quad (8)$$

$\omega^o_{m,j}$ is the reaction rate of each elementary reaction. Also, the term $\dot{\omega}_T$ in Eq. (3) represents the heat release from chemical reactions and is estimated as $\dot{\omega}_T = -\sum_{m=1}^{M} \dot{\omega}_m \Delta h^o_{f,m}$. $\Delta h^o_{f,m}$ is the formation enthalpy of $m$-th species.

## *2.2 Liquid phase*

The Lagrangian method is used to model the dispersed liquid phase, which is composed of a large number of spherical droplets [25]. The interactions between the droplets are neglected because we only study the dilute water sprays with the initial droplet volume fraction being generally less than 1‰ [26]. Droplet break-up is not considered due to the fine or ultra-fine droplets and will be studied in a future study. Therefore, the governing equations of mass, momentum and energy for a single droplet are

$$\frac{dm_d}{dt} = -\dot{m}_d, \tag{9}$$

$$\frac{d\mathbf{u}_d}{dt} = \frac{\mathbf{F}_d}{m_d}, \tag{10}$$

$$c_{p,d}\frac{dT_d}{dt} = \frac{\dot{Q}_c + \dot{Q}_{lat}}{m_d}, \tag{11}$$

where $m_d = \pi\rho_d d_d^3/6$ is the mass of a single droplet, and $\rho_d$ and $d_d$ are the droplet material density and diameter, respectively. $\mathbf{u}_d$ is the droplet velocity vector, $\mathbf{F}_d$ is the force exerted on the droplet. $c_{p,d}$ is the droplet heat capacity at constant pressure, and $T_d$ is the droplet temperature. In this work, both $\rho_d$ and $c_{p,d}$ are dependent on the droplet temperature $T_d$ [27], i.e.

$$\rho_d(T_d) = \frac{a_1}{a_2^{1+(1-T_d/a_3)^{a_4}}}, \tag{12}$$

$$c_{p,d}(T_d) = \frac{b_1^2}{\tau} + b_2 - \tau\left\{2.0b_1b_3 + \tau\left\{b_1b_4 + \tau\left[\frac{1}{3}b_3^2 + \tau\left(\frac{1}{2}b_3b_4 + \frac{1}{5}\tau b_4^2\right)\right]\right\}\right\}, \tag{13}$$

where $a_i$ and $b_i$ are constants and can be found from Ref. [27]. In Eq. (13), $\tau = 1.0 - min(T_d, T_c)/T_c$, where $T_c$ is the critical temperature.

The evaporation rate, $\dot{m}_d$, in Eq. (9) is modelled through

$$\dot{m}_d = -\dot{m}_f A_d, \tag{14}$$

where $A_d$ is the surface area of a single droplet. The vapor mass flux from the droplet into the gas phase, $\dot{m}_f$, is calculated as [28,29]

$$\dot{m}_f = k_c W_d (c_s - c_g). \tag{15}$$

Its accuracy has been confirmed through validations against the experimental data of single droplet evaporation [30]. $W_d$ is the molecular weight of the vapor. $c_s$ is the vapor mass concentration at the droplet surface, i.e.

$$c_S = \frac{p_{sat}}{R_g T_f}, \tag{16}$$

where $p_{Sat}$ is the saturation pressure and is obtained under the assumption that the vapor pressure at the droplet surface is equal to that of the gas phase. The droplet surface temperature is calculated from $T_f = (T + 2T_d)/3$ [29]. In Eq. (15), the vapor concentration in the surrounding gas, $c_g$, is obtained from

$$c_g = \frac{p x_i}{R_g T_f}, \tag{17}$$

where $x_i$ is the fuel vapor mole fraction in the surrounding gas. The mass transfer coefficient, $k_c$, in Eq. (15) is calculated from the Sherwood number $Sh$ [31], i.e.,

$$Sh_{ab} = \frac{k_c d_d}{D_f} = 2.0 + 0.6 Re_d^{1/2} Sc^{1/3}, \tag{18}$$

where $Sc$ is the Schmidt number of gas phase. The droplet Reynolds number in Eq. (18), $Re_d$, is defined based on the interphase velocity difference

$$Re_d \equiv \frac{\rho_d d_d |\mathbf{u}_d - \mathbf{u}|}{\mu}. \tag{19}$$

Moreover, $D_f$ in Eq. (18) is the vapor mass diffusivity in the gas phase [32]

$$D_f = 10^{-3} \frac{T_s^{1.75}}{p_s} \sqrt{\frac{1}{W_d} + \frac{1}{W_m}} / \left(V_1^{1/3} + V_2^{1/3}\right)^2, \tag{20}$$

where $V_1$ and $V_2$ are constants [33].

Since the ratio of gas density to the water droplet material density is well below one, the Basset force, history force and gravity force are not considered [34]. Only the Stokes drag $\mathbf{F}_d$ is considered in Eq. (10) and modelled as (assuming that the droplet is spherical) [35]

$$\mathbf{F}_d = \frac{18\mu}{\rho_d d_d^2}\frac{C_d Re_d}{24} m_d(\mathbf{u} - \mathbf{u}_d). \tag{21}$$

The drag coefficient in Eq. (21), $C_d$, is estimated as [35]

$$C_d = \begin{cases} 0.424, & if \;\; Re_d \geq 1000, \\ \frac{24}{Re_d}\left(1 + \frac{1}{6} Re_d^{2/3}\right), & if \;\; Re_d < 1000. \end{cases} \tag{22}$$

It has been shown by Cheatham and Kailasanath [36] that Eq. (22) can correctly predict the velocity distributions of a flow field with shock waves, and has the comparable accuracies to other models for drag coefficients [37–39].

The convective heat transfer rate $\dot{Q}_c$ in Eq. (11) is calculated from

$$\dot{Q}_c = h_c A_d (T - T_d). \tag{23}$$

Here $h_c$ is the convective heat transfer coefficient, and computed using the correlation by Ranz and Marshall [31], i.e.,

$$Nu = h_c \frac{d_d}{k} = 2.0 + 0.6 Re_d^{1/2} Pr^{1/3}, \tag{24}$$

where $Nu$ and $Pr$ are the Nusselt and Prandtl numbers of gas phase, respectively. In addition, the heat transfer associated with droplet evaporation, $\dot{Q}_{lat}$ in Eq. (11), is

$$\dot{Q}_{lat} = -\dot{m}_d h(T_d), \tag{25}$$

where $h(T_d)$ is the heat of vaporization at the droplet temperature $T_d$.

Two-way coupling between the gas and liquid phases is enforced in this work. The corresponding terms, $S_{mass}$, $\mathbf{S}_{mom}$, $S_{energy}$ and $S_{species,m}$ in Eqs. (1) – (4), are calculated based on the contributions from each droplet in the CFD cells, which read

$$S_{mass} = \frac{1}{V_c}\sum\nolimits_1^{N_d} \dot{m}_d, \tag{26}$$

$$\mathbf{S}_{mom} = -\frac{1}{V_c}\sum\nolimits_1^{N_d} (-\dot{m}_d \mathbf{u}_d + \mathbf{F}_d), \tag{27}$$

$$S_{energy} = -\frac{1}{V_c}\sum_{1}^{N_d}(\dot{Q}_c + \dot{Q}_{lat}) \, , \quad (28)$$

$$S_{species,m} = \begin{cases} S_{mass} & for\ H_2O\ species \\ 0 & for\ other\ species. \end{cases} \quad (29)$$

In the following, $S_{mass}$, $\mathbf{S}_{mom}$ and $S_{energy}$ are termed as mass, momentum, energy transfer rates, respectively. Here $V_c$ is the CFD cell volume and $N_d$ is the droplet number in one cell. In Eq. (27), $-\dot{m}_d\mathbf{u}_d$ is the momentum transfer due to droplet evaporation.

### 2.3 *Computational method*

The gas and liquid phase governing equations are solved by a compressible two-phase reacting flow solver, *RYrhoCentralFoam* [40], which is customized from the fully compressible non-reacting flow solver *rhoCentralFoam* in OpenFOAM 6.0 [41]. *rhoCentralFoam* is extensively validated using the Sod's problem, forward step, supersonic jet and shock−vortex interaction [42,43]. Moreover, *RYrhoCentralFoam* has been extensively validated and verified for detonation problems in gaseous and gas−droplet two-phase flows, and good agreements are achieved about detonation frontal structure, cell size, propagation speed and gas−liquid two-phase coupling [44]. It has been successfully applied for various detonation problems [45–49].

Second-order implicit backward method is employed for temporal discretization and the time step is about $1\times10^{-11}$s (maximum Courant number < 0.1). The KNP (i.e. Kurganov, Noelle and Petrova[50]) scheme with van Leer limiter is used for MUSCL-type reconstructions of the convective fluxes in momentum equation. Total Variation Diminishing (TVD) scheme is used for the convective terms in energy and species mass fraction equations. Also, second-order central differencing scheme is applied for the diffusion terms in Eqs. (2)−(4). The hydrogen mechanism with 9 species and 19 reactions [51] is used, which is validated against the measured ignition delay and detonation cell size [44].

For the liquid phase, the droplets are tracked based on their barycentric coordinates. The equations, i.e., Eqs. (9)−(11), are solved by first-order implicit Euler method. Meanwhile, the gas properties at the droplet location (e.g., the gas velocity in Eq. 19 and temperature in Eq. 22) are calculated based on linear interpolation. More detailed information about the numerical methods for gas and liquid phases can be found in Ref. [40].

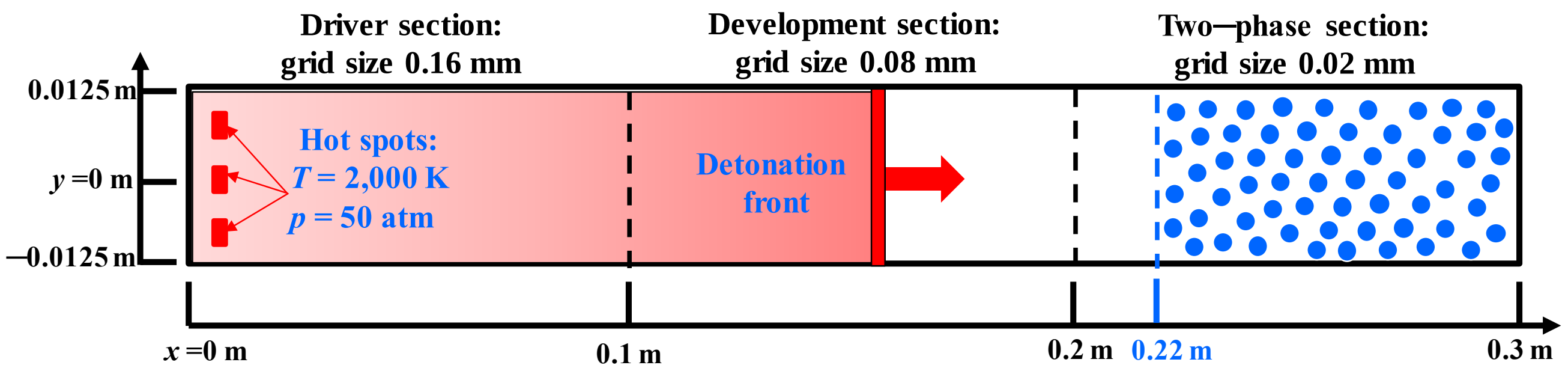

Figure 2 Schematic of the computational domain. Blue dots: water droplets.

## 3. Physical model

The computational domain is shown in Fig. 2. The length ($x$-direction) and width ($y$-direction) are 0.3 m and 0.025 m, respectively. It includes driver, development, and two-phase sections, as marked in Fig. 2. They are initially filled with stoichiometric $H_2$/air premixture, with temperature and pressure being $T_0$ = 300 K and $p_0$ = 50 kPa, respectively. Uniform Cartesian cells are used to discretize the domain in Fig. 2, and the mesh cell size transitions from 160 μm in the driver section, 80 μm in the development section, to 20 μm in the two-phase section. The total cell numbers in the three sections are 102,400, 409,600, and 6,250,000, respectively. The Half-Reaction Length (HRL) estimated from the purely gaseous ZND structure of $H_2$/air detonation is 309 μm. Therefore, the resolution in the two-phase section, where our analysis is focused, is approximately 15 cells per HRL. The total length of the driver and development sections is about 647 HRL, and hence sufficient to minimize the detonation initiation effects before the Detonation Wave (DW) is transmitted into the two-phase section [52]. A halved mesh resolution (10 μm) is also tested for the last section and it is shown (see supplementary material) that predicted detonation cell sizes are close to those with 20 μm.

The DW is initiated by three vertically placed hot spots (2,000 K and 50 atm) at the left end (see Fig. 2), and the interactions between the shock waves can quickly lead to the detonation frontal cellularity. The purely gas result when the DW lies at $x$ = 0.2 m (slightly before the two-phase section) is used as the initial field for all the two-phase simulations. The upper and lower boundaries of the domain in Fig. 2 are assumed to be periodic. For the left boundary ($x$ = 0), the non-reflective condition is enforced for the pressure, while the zero gradient condition for other quantities [53]. Since the gas before the detonation wave is static, the boundary condition at $x$ = 0.3 m is not relevant and in our simulations zero gradient conditions are assumed [54].

The monodispersed spherical water droplets are uniformly distributed in the two-phase section (i.e., $x$ = 0.22−0.3 m). The initial water droplet diameters $d_d^0$ range from 2.5 to 15 μm, which roughly correspond to the dominant sizes of the water droplets from ultrasonic mist generators [55]. Although water droplet polydispersity is ubiquitous in practical scenarios [14,16], however, monodispersed droplets are helpful for pinpointing the droplet size effects and the polydispersity effects merit a separate study. The mass loading $z$ = 0.031 − 0.93 will be studied in this work, corresponding to water apparent density $\varrho$ of 0.013 − 0.391. Note that $z$ (or $\varrho$) is estimated as the ratio of the total water mass to the mass (or volume) of the gaseous mixture within the droplet-containing region [26]. The initial temperature, material density and isobaric heat capacity of the water droplets are 300 K, 997 kg/m$^3$ and 4,187 J/kg·K, respectively. Besides, the water droplets are assumed to be initially stationary (i.e., $\mathbf{u}_d = 0$), which is reasonable due to typically small terminal velocities of fine water droplets [9,10].

## 4. Results

### 4.1 Detonation extinction diagram

A series of $H_2$/air detonations in water mists is simulated in this work. They are parameterized by a range of initial droplet diameter $d_d^0$ and mass loading $z$ (or apparent density $\varrho$), which are summarized in Fig. 3. The water vapour mass fraction in the mixture ($Y_{H2O}$) is also marked for each mass loading, assuming that the water sprays are fully gasified. Generally, propagation of hydrogen/air detonation is considerably influenced by both water droplet size and mass loading. Figure 3 can be divided into two regimes. Specifically, the cases left to the dashed line correspond to successful detonation propagation in the two-phase section. For these cases, the lower water mass loading, the stronger the detonative combustion. This is featured by higher averaged Heat Release Rate (HRR), which is time-averaged volume-integrated HRR in the two-phase section. Some of them are selected to be further simulated with extended length (i.e., $x$ = 0.22−0.4 m) of the two-phase section and it is shown that the DW can still propagate beyond $x$ = 0.3 m.

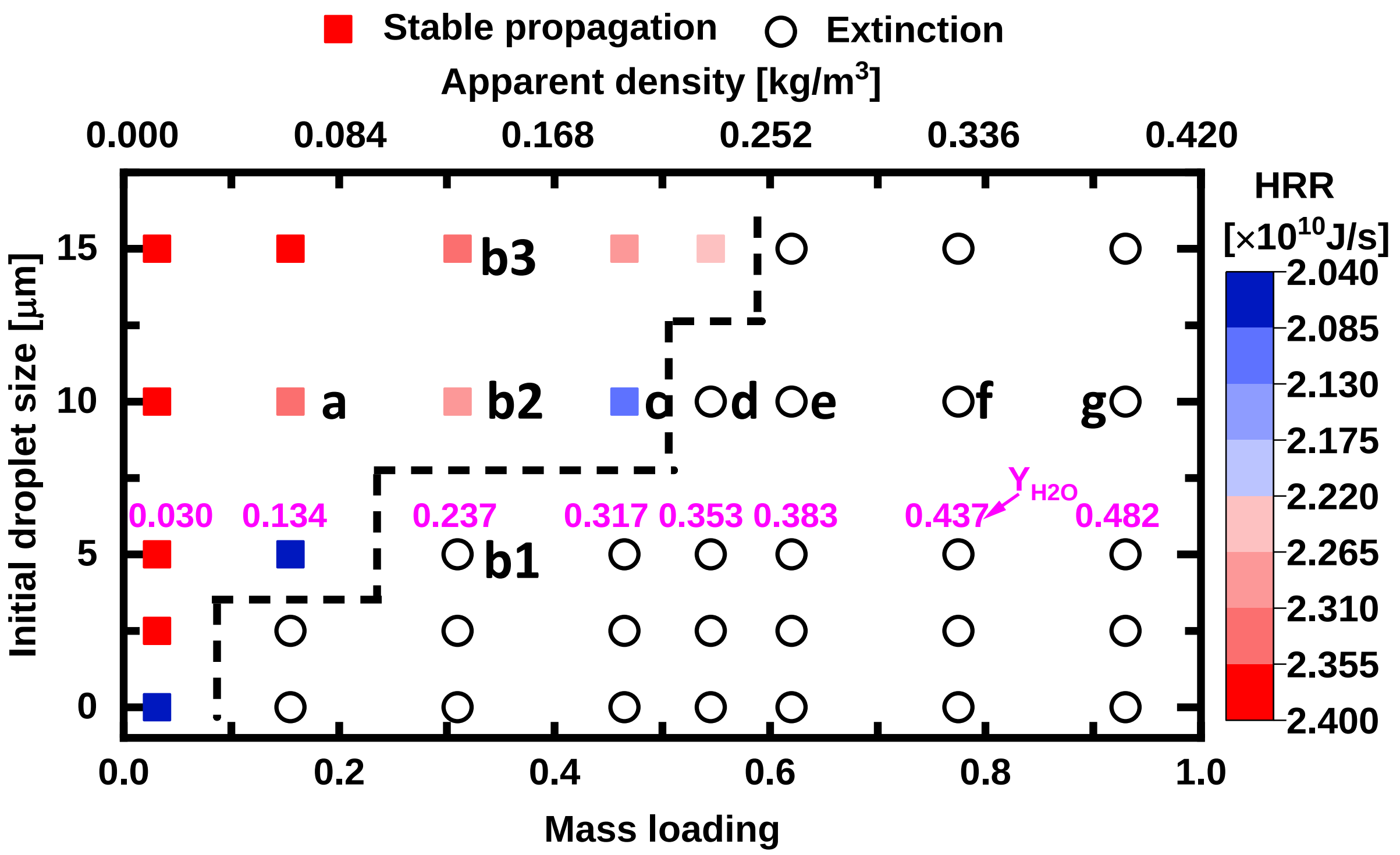

Figure 3 Diagram of two-phase detonation propagation and extinction. $Y_{H2O}$: water vapor when all the droplets are vaporized.

Moreover, the rest cases with open symbols correspond to detonation extinction, characterized by the ultimate decoupling of the reaction and leading shock fronts after the DW's travel a finite distance in the two-phase section. When the mass loading $z$ is beyond 0.6 (correspondingly $\varrho$ > 0.252), detonation extinction always occurs, regardless of the droplet diameters (i.e., 2.5−15 μm). However, when $z$ < 0.6, the droplet diameter effects appear. Specifically, for a fixed mass loading, the DW's are quenched with relatively small $d_d^0$, and the critical diameter for detonation extinction decreases with smaller $z$. The purely gas cases ($d_d^0$ = 0 in Fig. 1) are also run when the liquid water is fully vaporized (with $T_0$ = 300 K and $p_0$ = 50 kPa). One can see that, except with $z$ = 0.031, the detonations are quenched for all the loadings. This indicates that the $H_2$/air mixtures with the above diluent concentrations cannot support detonation propagation. This tendency is also seen with ultra-fine droplets (e.g., 2.5 μm), implying that the significant kinetic contributions of the water vapour from their evaporation.

Nine representative two-phase cases will be discussed in detail in the following, which are cases a−g in Table 1. Specifically, cases a, b2, and c−g are selected to study the water mass loading effects with a fixed droplet diameter, i.e., $d_d^0$ = 10 μm. Furthermore, cases b1, b2, and b3 with the same mass loading $z$ = 0.31 are used to examine the influence of initial droplet size on propagation and extinction of detonation waves. The cases in Table 1 are also marked in Fig. 3. Moreover, the purely gaseous detonation in stoichiometric $H_2$/air mixture, case h, is also simulated as a reference case.

Table 1 Selected simulated cases

| **Case** | | | **Mass loading** $z$ | **Diameter** $d_d^0$/**μm** |
|---|---|---|---|---|
| Two-phase mixture ($H_2$/air + water droplets) | a | | 0.155 | 10 |
| | b | 1 | 0.31 | 5 |
| | | 2 | | 10 |
| | | 3 | | 15 |
| | c | | 0.465 | 10 |
| | d | | 0.545 | |
| | e | | 0.62 | |
| | f | | 0.775 | |
| | g | | 0.93 | |
| Purely gaseous mixture ($H_2$/air) | h | | – | – |

**4.2 General reaction zone structure of detonation in water mists**

Figure 4 shows the instantaneous pressure, temperature, and heat release rate from case b2 ($z$ = 0.31 and $d_d^0$ = 10 μm). It is shown that the DW can stably propagate in the two-phase section with multiple transverse waves and detonation heads. Due to droplet evaporation, the gas temperature is reduced to about 1,600 K at around $x$ = 0.224 – 0.26 m, different from purely gaseous detonation [56,57]. Heat release mainly occurs behind the leading Shock Front (SF) and Transverse Wave (TW). Figure 5 shows the enlarged view of the detonation front cellular structure corresponding to the dashed box in Fig. 4. One can see from Fig. 5(a) that the Mach Stems (MS), Incident Waves (IW), TW, and primary / secondary Triple Points (TP1 / TP2) are captured. As marked in Fig. 5(b), the primary and secondary Jet Flows (JF1 and JF2) are also predicted, which are respectively generated through Richtmyer–Meshkov instability [58] and collision between the neighboring triple points. Pockets of the unreacted gas escape from the main detonation front after the collision of two triple points, as marked in Figs. 5(b) and 5(c). Moreover, the temperature and HRR behind the MS are much higher than that behind the IW. Decoupling between the Reaction Front (RF) and IW can be seen, which further produces the double Mach structure [59], as indicated in Fig. 5(c) and the inset. The detailed thermochemical states behind the MS and IW will be further interpreted with CEMA in Section 5.2.

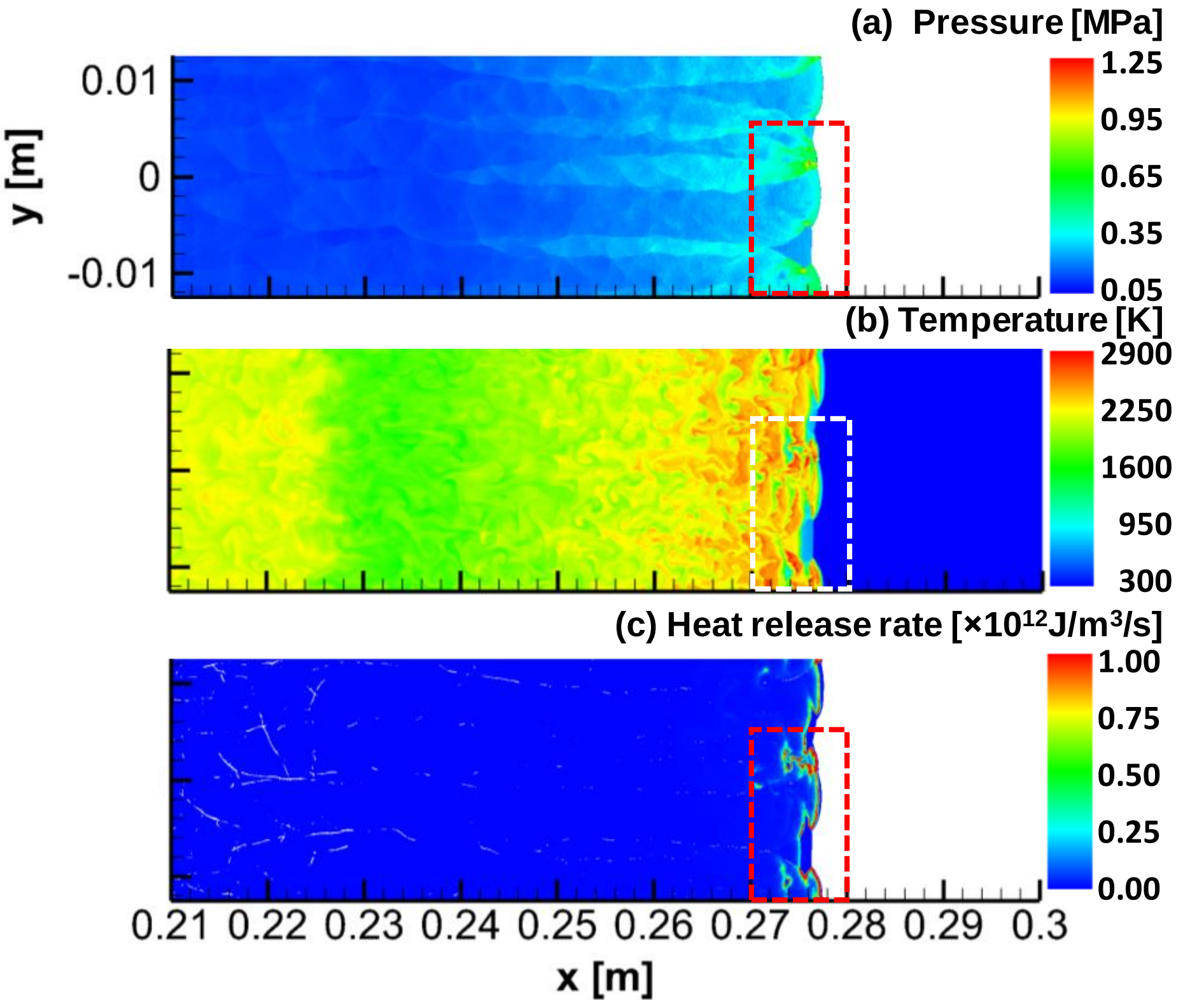

Figure 4 Distributions of (a) pressure, (b) temperature, and (c) heat release rate in the two-phase section. Results from case b2: $z$ = 0.31 and $d_d^0$ = 10 μm.

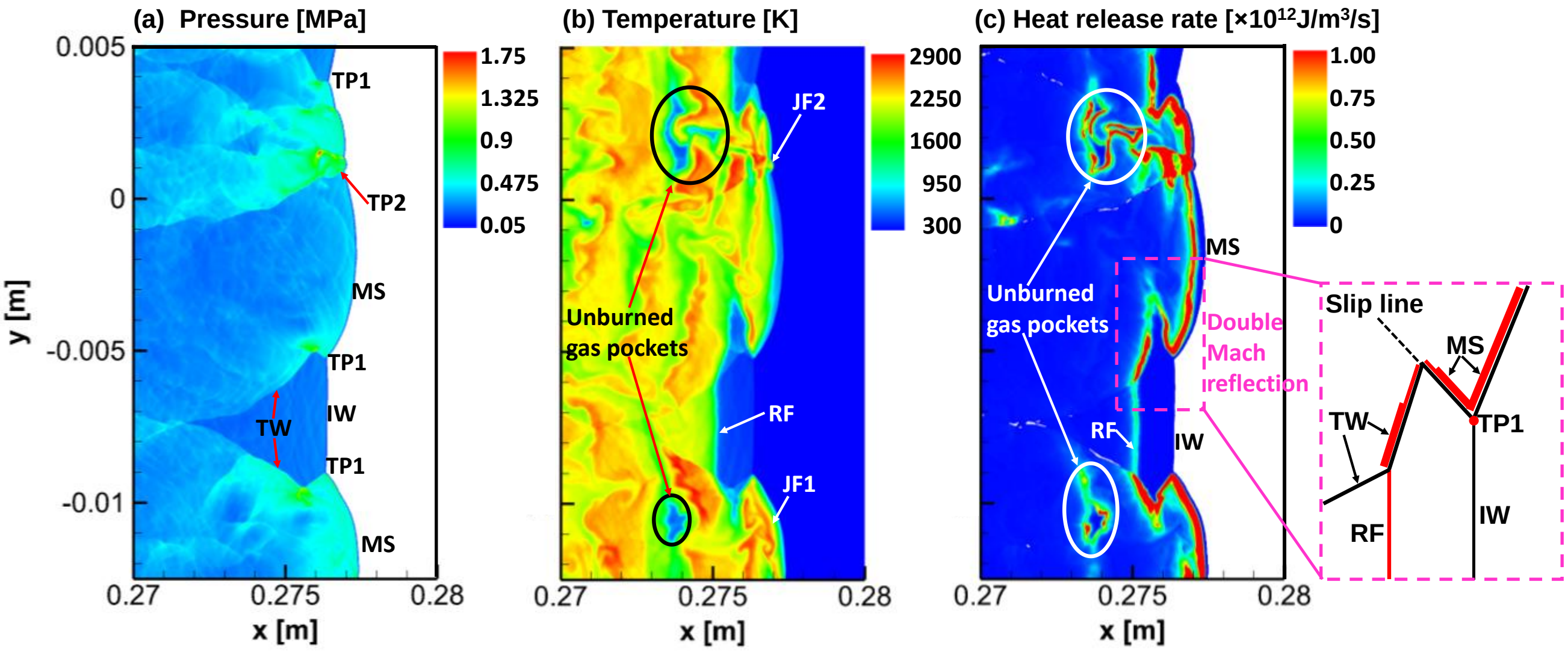

Figure 5 Close-up view of detonation frontal structure in the dashed box in Fig. 4: (a) pressure, (b) temperature, and (c) heat release rate.

Figure 6 shows the distributions of instantaneous diameter $d_d$, evaporation rate $\dot{m}_d$ and temperature $T_d$ for Lagrangian water droplets, corresponding to the same instant as in Figs. 4 and 5. The details inside the dashed box are shown in Fig. 7. To describe the droplet evaporating zone, two characteristic locations are denoted in Fig. 6, i.e., Evaporation Onset Front (EOF) and End of Two-phase Section (ETS). In our analysis, EOF corresponds to a location where the local droplet evaporation rate is larger than $1\times 10^{-9}$ kg/s, whilst the End of Two-phase Section (ETS) is the contact surface between the purely gaseous and two-phase mixtures.

As shown in Fig. 6(a), the droplet diameter before the DW remain unchanged (10 μm), due to limited evaporation. The droplet diameter slightly increases by roughly 1% immediately behind the DW due to the droplet expansion (also shown in Fig. 7a). This is caused by elevated droplet temperature (Fig. 6c), due to slight droplet density reduction (see Eq. 12). Since the mass of the individual droplet is almost not changed, the droplet volume (hence diameter) slightly increases. Although the gas temperature immediately behind the MS is much higher than that behind the IW, nevertheless, the droplet heating takes a finitely long distance behind both shock waves, as demonstrated in Fig. 7(c). In this case, the distance between the DW and EOF is about 5 mm, which is smaller than the counterpart result (about 10 mm) in $C_2H_4$/air detonation in water sprays [17]. This is reasonable because larger droplet diameter (15.9 μm) is considered in Ref. [17] and hence longer droplet heating period is expected. The droplet temperature is increased to 440 K around the EOF and strong droplet evaporation then occurs. Note that the heated droplets mainly exist behind the Reaction Front (RF in Fig. 7) where pronounced heat release occurs. The evaporation continues in the entire post-detonation area (see Fig. 6b), although the evaporation rate slightly decreases further downstream. This may be associated with the lower gas and water droplet temperatures, as demonstrated in Figs. 4(b) and 6(c).

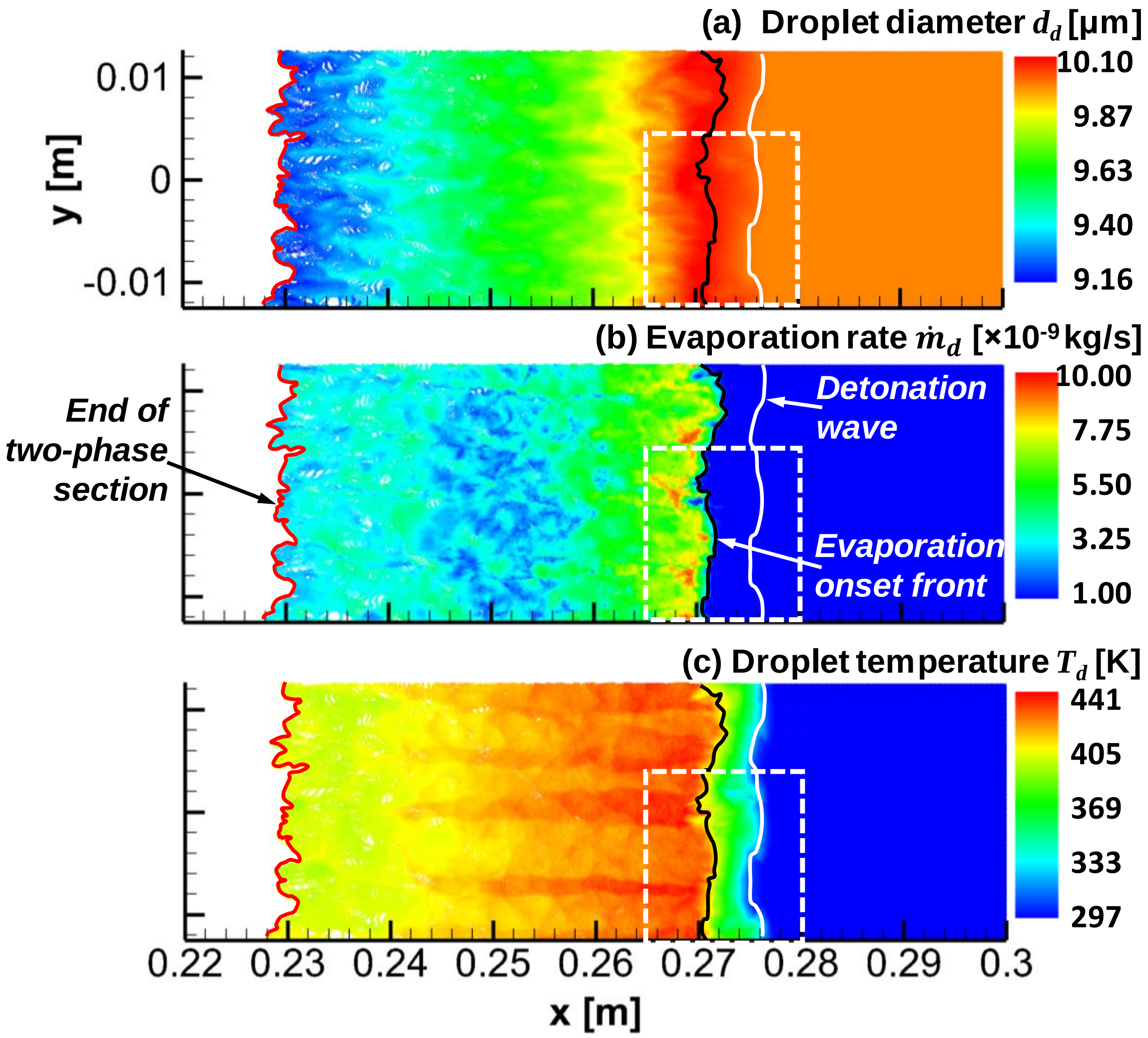

Figure 6 Distributions of Lagrangian water droplets colored with instantaneous (a) diameter, (b) evaporation rate, and (c) temperature.

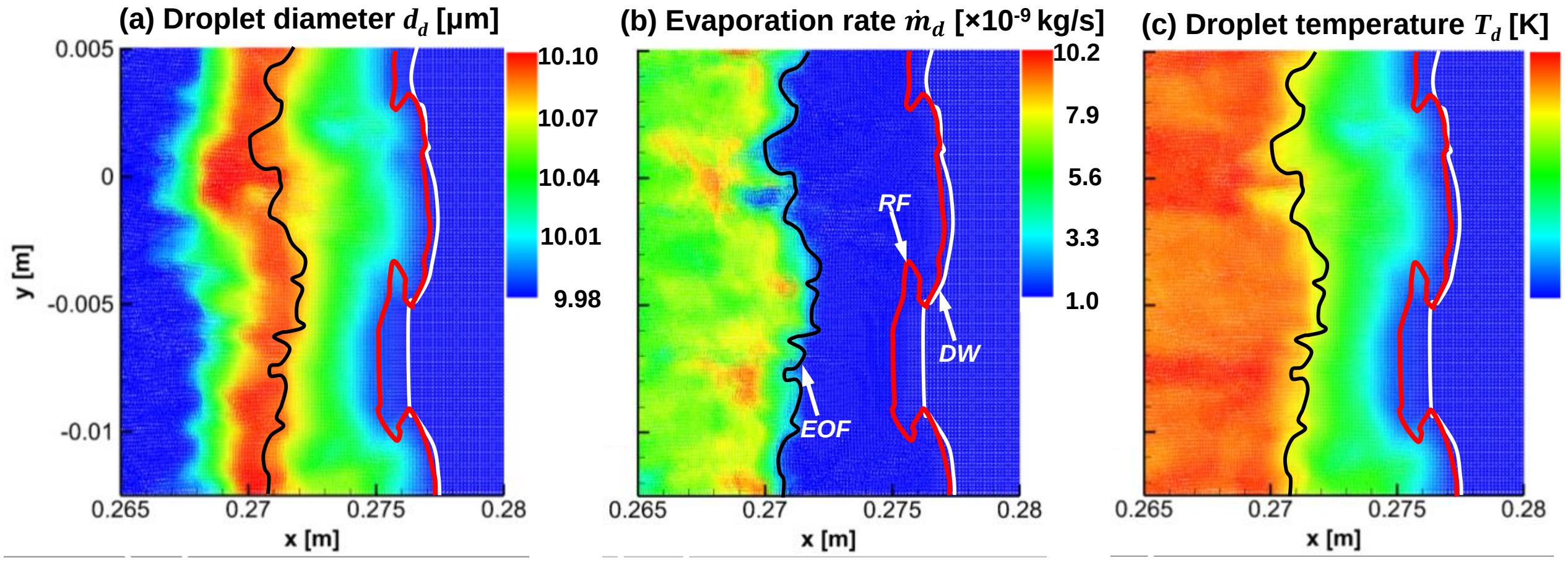

Figure 7 Close-up view of the Lagrangian droplets in the dashed box in Fig. 6: (a) diameter, (b) evaporation rate, and (c) temperature.

### 4.3 Droplet mass loading effects

To demonstrate the influences of water spray mass loading on $H_2$/air detonations, numerical soot foils are demonstrated in Fig. 8, which are recorded from the trajectory of maximum pressure location when the DW propagates in the two-phase section. It is known that these tracks are closely associated with the triple points on the detonation front and the cell apexes are the loci of triple point collision [60]. Gaseous detonation, case h, is also included in Fig. 8(h) for comparison. It can be observed that the presence of water droplets considerably changes the peak pressure trajectories of stoichiometric $H_2$/air detonations. Additionally, in the upstream of the two-phase section, the cells of cases a−c increase with mass loading, i.e., 5.9, 7.3 and 8.9 mm, which are higher than that in case h (5.2 mm). This tendency is also observed in spray $C_2H_4$/air experiments by Jarsalé et al. [16]. Generally, the cell width is proportional to the induction length [61–63], which may increase due to stronger evaporating cooling and/or more water vapor dilution as $z$ increases. Moreover, this spatial variation of the cells is not reported by Watanabe et al. [17] in their simulations of the $C_2H_4$/air detonations in water sprays. It may be related to the higher reduced activation energy of hydrogen/air mixture. Furthermore, in cases a−c, the leading shock wave propagation speed within one cell varies from 0.4 and 1.4 of the Chapman–Jouguet velocity, and these deviations are close to those in gaseous detonations [61,64,65].

Two striking features emerge due to the movement of triple points (indicated by the white arrows in Fig. 8): merged trajectory in Fig. 8(a) and re-amplification of the new triple points in Figs. 8(b) and 8(c). The latter is also observed in the simulations of gaseous detonation propagation in channels with porous walls and is attributed to the interactions of transverse wave with different intensities [66]. However, transition into the single head mode is not seen in our results, different from the observations from ethylene/air detonation in water sprays [16].

For the cases d, e, f, and g, the DW propagates a distance in the two-phase section, and then the leading shock front and reaction front decouples. This is characterized by the quickly reduced peak pressure and therefore faded trajectories in Figs. 8(d) – 8(g). Moreover, when the mass loading increases, the detonation extinction occurs earlier.

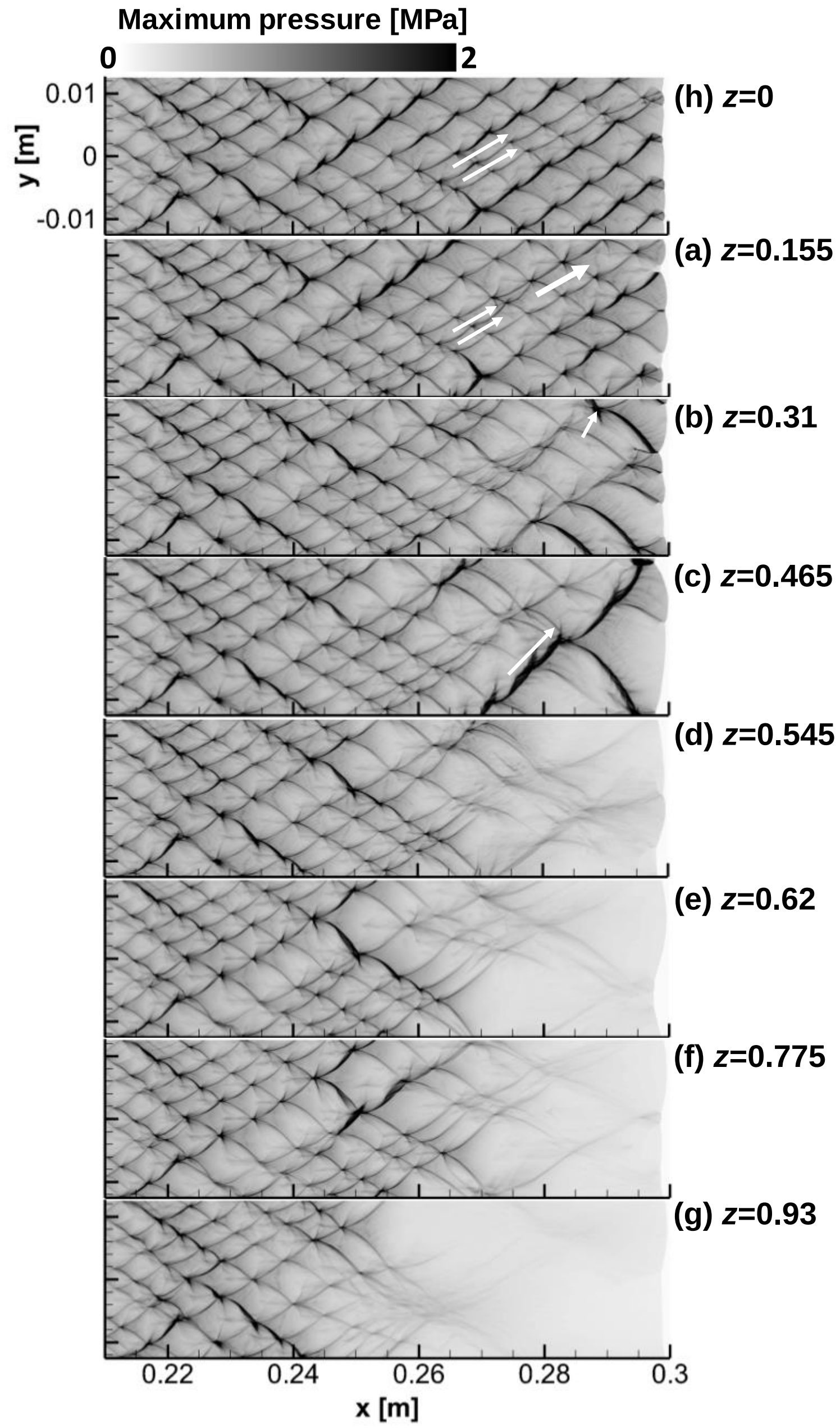

Figure 8 Numerical smoked foils of (h) gaseous and (a−g) two-phase detonations with different mass loadings.

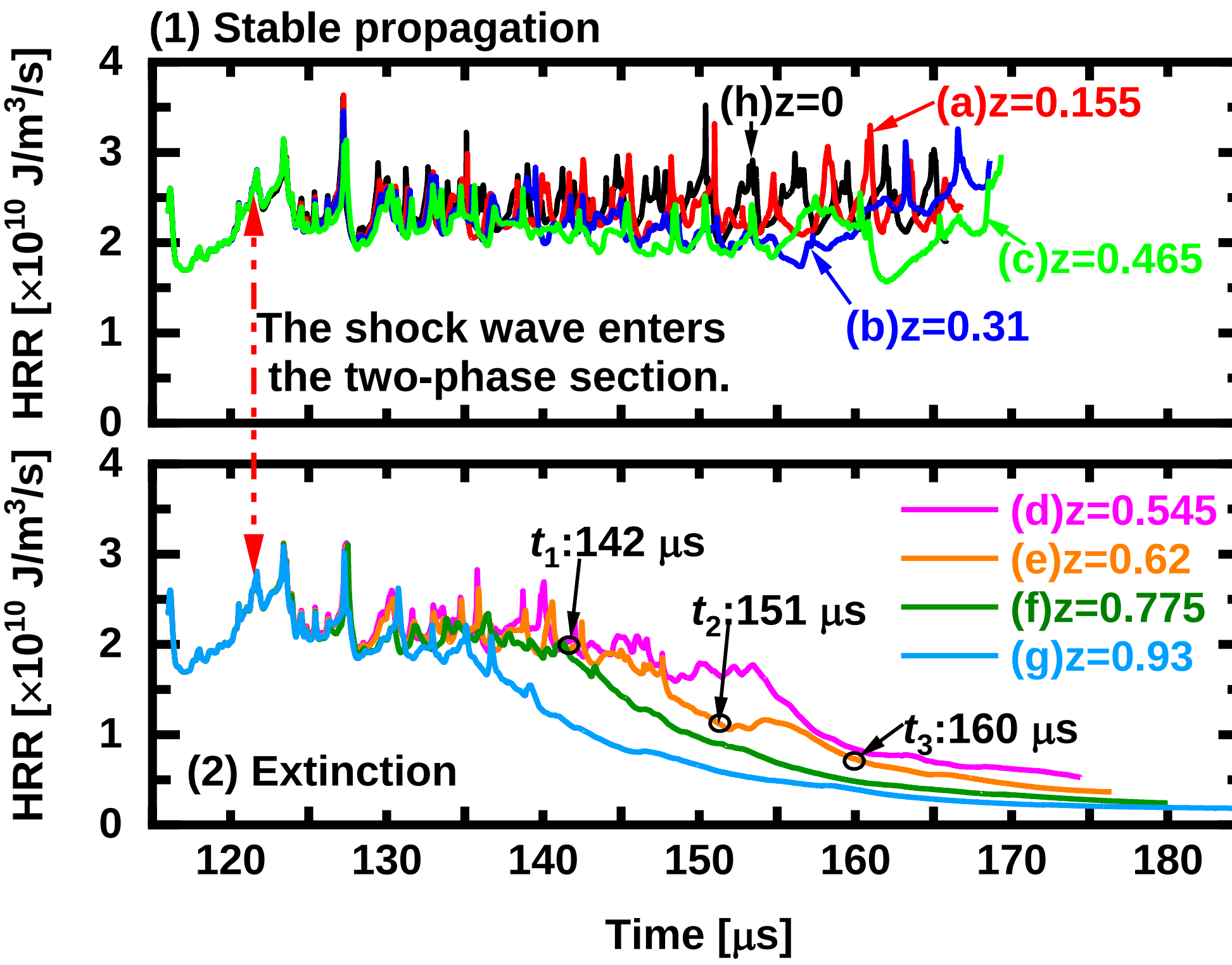

Figure 9 Temporal evolution of volume-averaged heat release rate in pure gas and two-phase detonations with different mass loadings ($d_d^0$ = 10 μm).

Figure 9 further quantifies the detonation extinction process through the time evolutions of volume-averaged HRR in the corresponding cases in Fig. 8. The detonation wave enters the two-phase section at about 122 μs. For cases a and b, the averaged HRR fluctuates regularly, and is similar to that of pure gas case h. The periodic variations of heat release arise from the collisions between transverse waves and triple points [67]. However, in case c, when $t$ > 160 μs, the HRR gradually increases, which is due to the collisions of the triple points, leading to a new re-amplified one. For the DW extinction cases (d−g), the heat release gradually decreases. This indicates the global detonation extinction, without any re-initiation.

### 4.4 Droplet size effects

Figure 10 shows the numerical soot foils of hydrogen/air detonations with initial water droplet diameters of 5, 10 and 15 μm (i.e., b1, b2 and b3, respectively, in Table 1). The water mass loading is fixed to be $z$ = 0.31. The reader is reminded that Fig. 10(b) is the same as that in Fig. 8(b). The DW is decoupled after propagating a distance in water sprays with $d_d^0$ = 5 μm. Therefore, the cellular structure in Fig. 10(a) is different from those in Figs. 10(b) and 10(c). For stable detonations, the average cell size in Figs. 10(b) is larger than that in Fig. 10(c) in the second half of the two-phase section. The addition of water droplet increases the autoignition delay time of the gaseous mixture because of vapour dilution and interphase heat transfer. At the same mass loading, smaller droplets indicate more water droplets, and therefore foregoing influences are more pronounced.

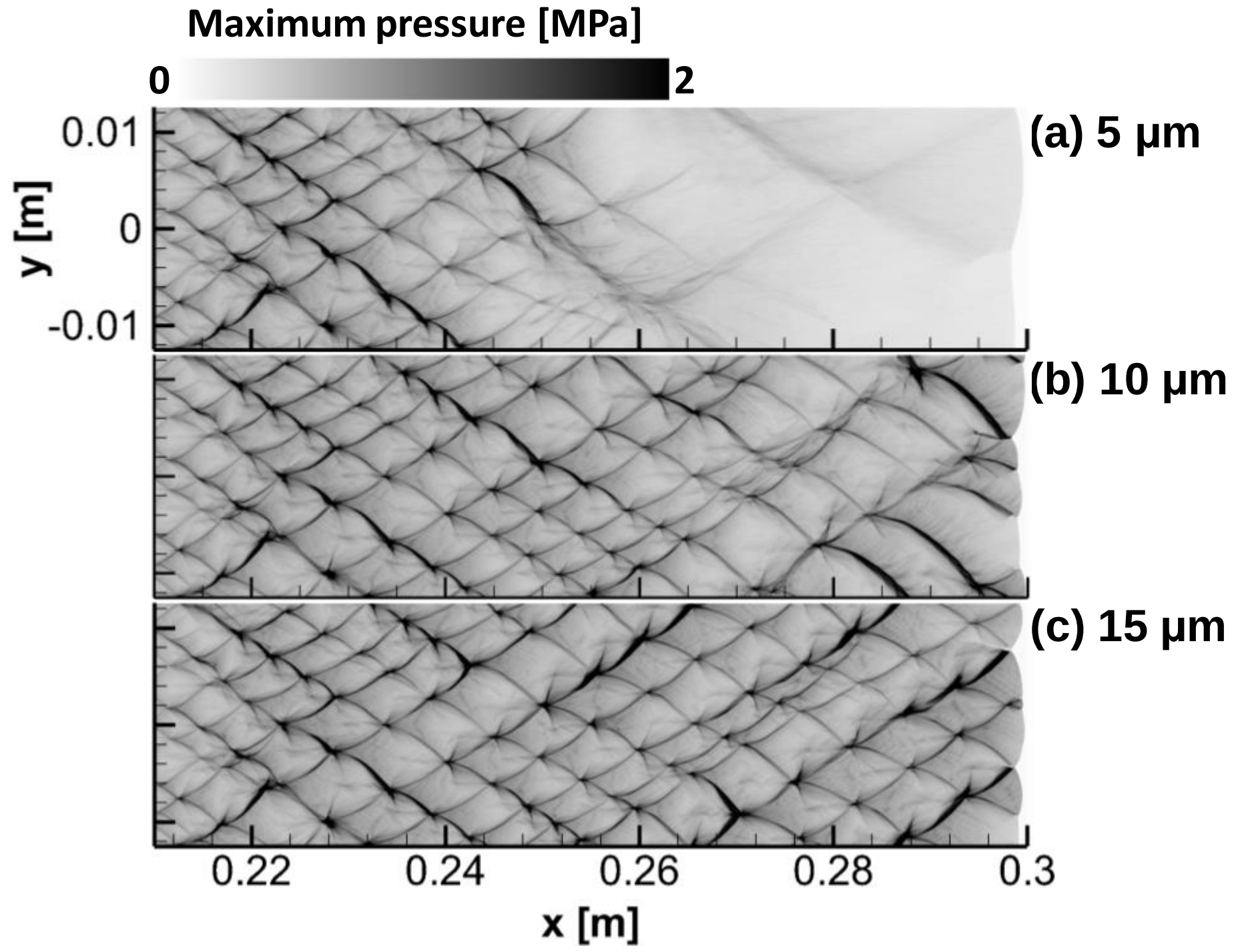

Figure 10 Numerical smoked foils of two-phase detonations with three droplet diameters: (a) 5 μm, (b) 10 μm and (c) 15 μm. *z* = 0.31.

Potted in Fig. 11 are the time evolutions of volume-averaged HRR from three cases in Fig. 10. It is observed that the HRR fluctuates around a constant with $d_d^0$ = 10 and 15 μm. Since triple point re-amplification is not seen when $d_d^0$ = 15 μm, elevation of HRR at a later stage of DW propagation does not occur at the after $t$ = 155 μs. However, when $d_d^0$ = 5 μm, the DW fails with continuously reduced HRR in Fig. 11.

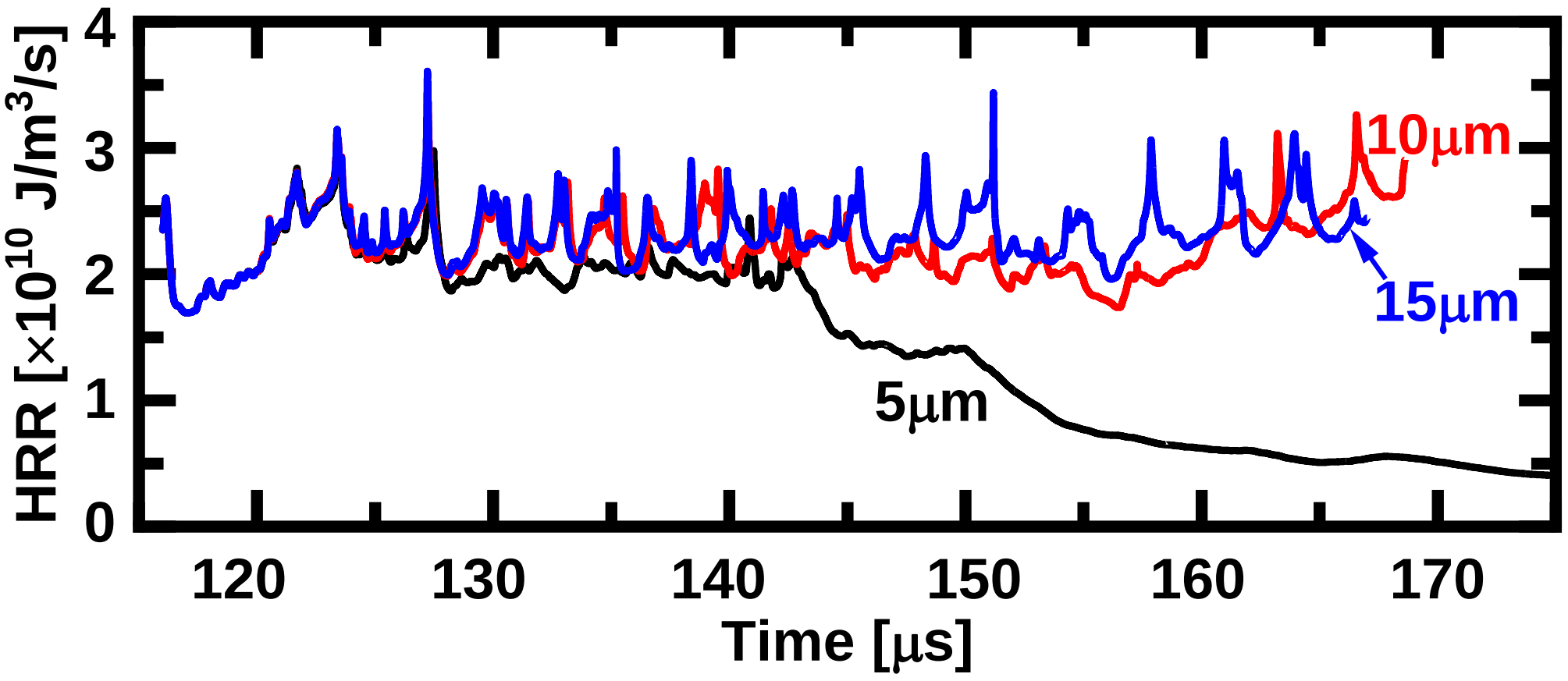

Figure 11 Temporal evolution of volume-averaged heat release rate with three droplet diameters: (a) 5 μm, (b) 10 μm and (c) 15 μm. $z = 0.31$.

## 5. Discussion

### 5.1 Detonation extinction transient

It can be seen from Figs. 3, 8 and 10 that when the water mass loading is large (or sprayed droplet size is small) hydrogen detonation extinction occurs. In this section, the unsteady detonation extinction process in case e will be further analysed. Figure 12 shows the time evolutions of gas temperature, water droplet evaporation rate, and water droplet temperature during the detonation extinction process. Note that Lagrangian droplets are visualized in Figs. 12(b) and 12(c). This unsteady process is also visualized with the animations (see supplementary material). The DW stably propagates at 142 μs. At 151 μs, partial extinction behind the leading SF can be seen, characterized by the increased distance between the SF and RF at most locations of the leading front. The SF and RF are completely decoupled at 160 μs. Meanwhile, one can see from Figs. 12(b) and 12(c) that generally the length of the shocked two-phase section (between SF and ETS) increases when the DW is gradually weakened. Due to the spatial delay of the RF relative to the SF, it takes longer for the water droplets to be heated and the evaporation becomes weaker behind the EOF. The two-phase heat exchange further weakens the deflagrative combustion in the shocked hydrogen/air mixtures.

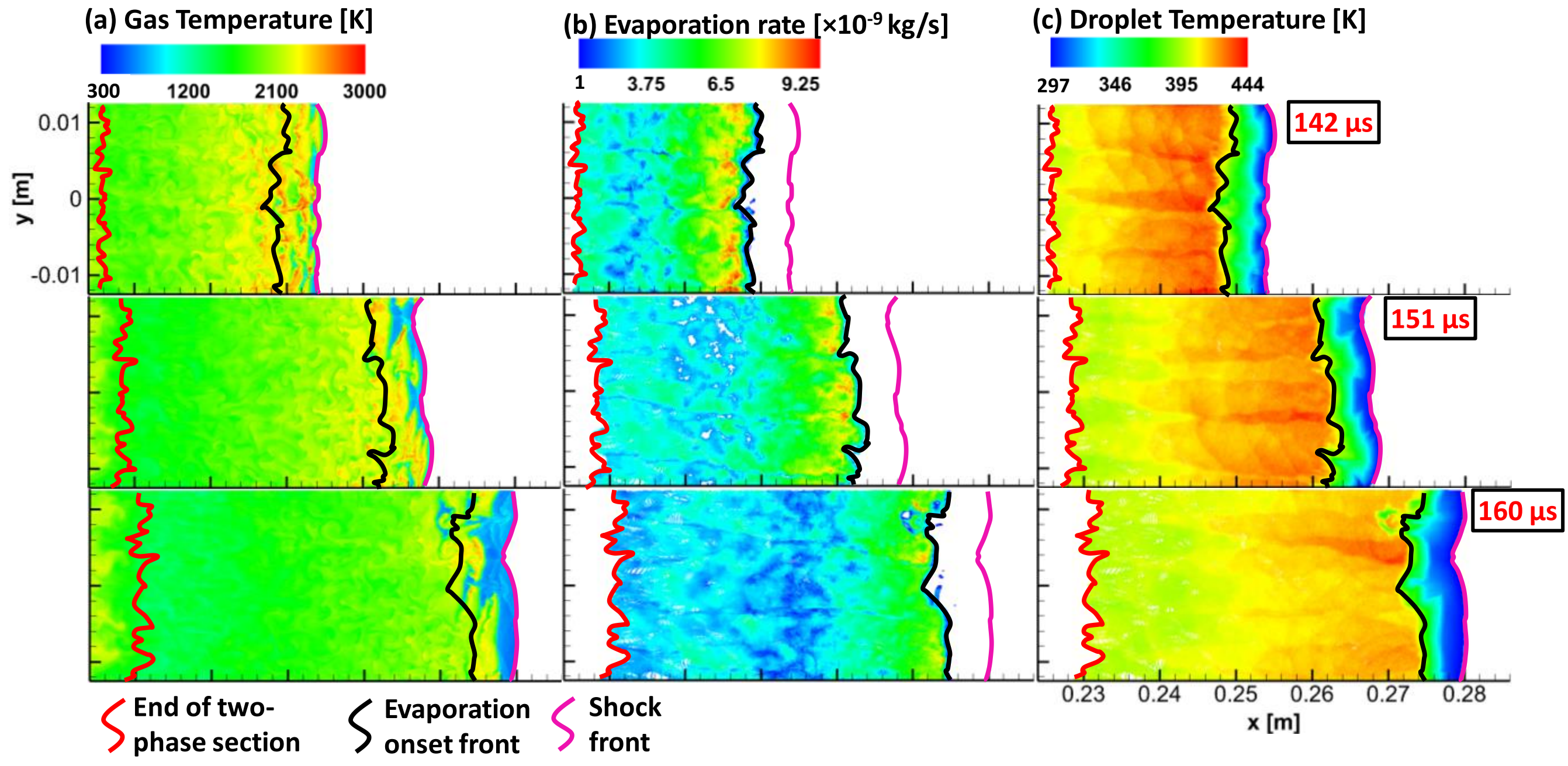

Figure 12 Evolution of the detonation front in three instants: (a) gas temperature, (b) droplet evaporation rate, and (c) droplet temperature. Results from case e with z = 0.62.

The evolutions of detailed frontal structure in the foregoing extinction process are further illustrated in Fig. 13. It is found that at 149 μs a series of Mach stems (e.g., MS0, MS1 and MS2) exist. As the two triple points of MS1 and MS2 move towards each other, their collision produces the third Mach stem (MS3) at 151 μs. However, no chemical reactions proceed at these two triple points and meanwhile their pressure superposition does not induce chemical reactions behind MS3. A jet flow, JF1, is generated behind MS3 at 152.5 μs. However, this cold jet does not initiate a detonation. When $t$ = 149−152.5 μs, MS0 is still followed by considerable heat release. At 154 μs, however, decoupling of reaction front and lower part of leading shock begins. A new Mach stem, MS4, is produced by the interaction of MS3 and the lower one. Finally, after $t$ = 157 μs, the RF is fully decoupled from the leading SF. Their distance gradually increases, and no detonation initiation is seen.

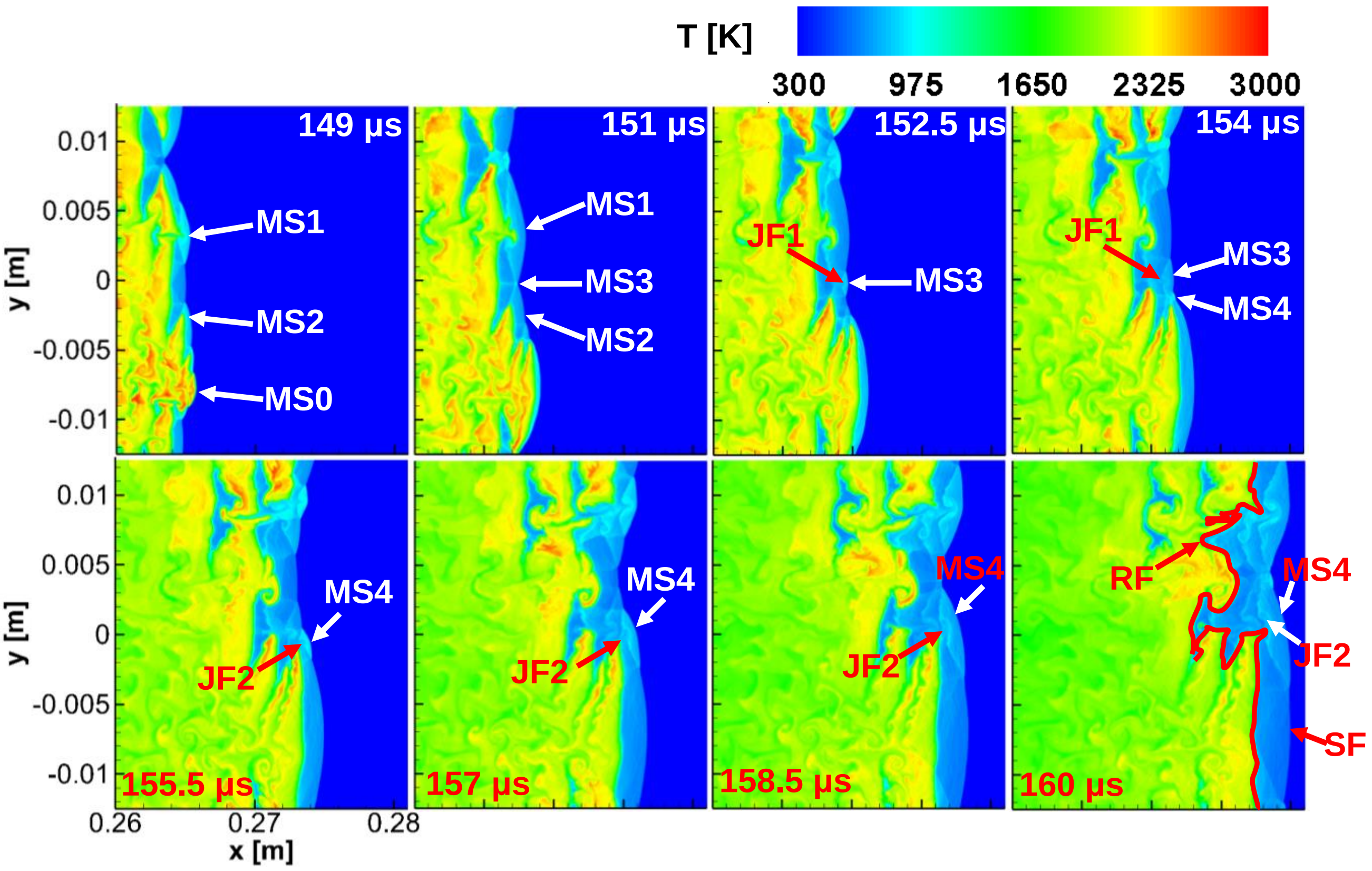

Figure 13 Evolution of gas temperature during a detonation extinction process.

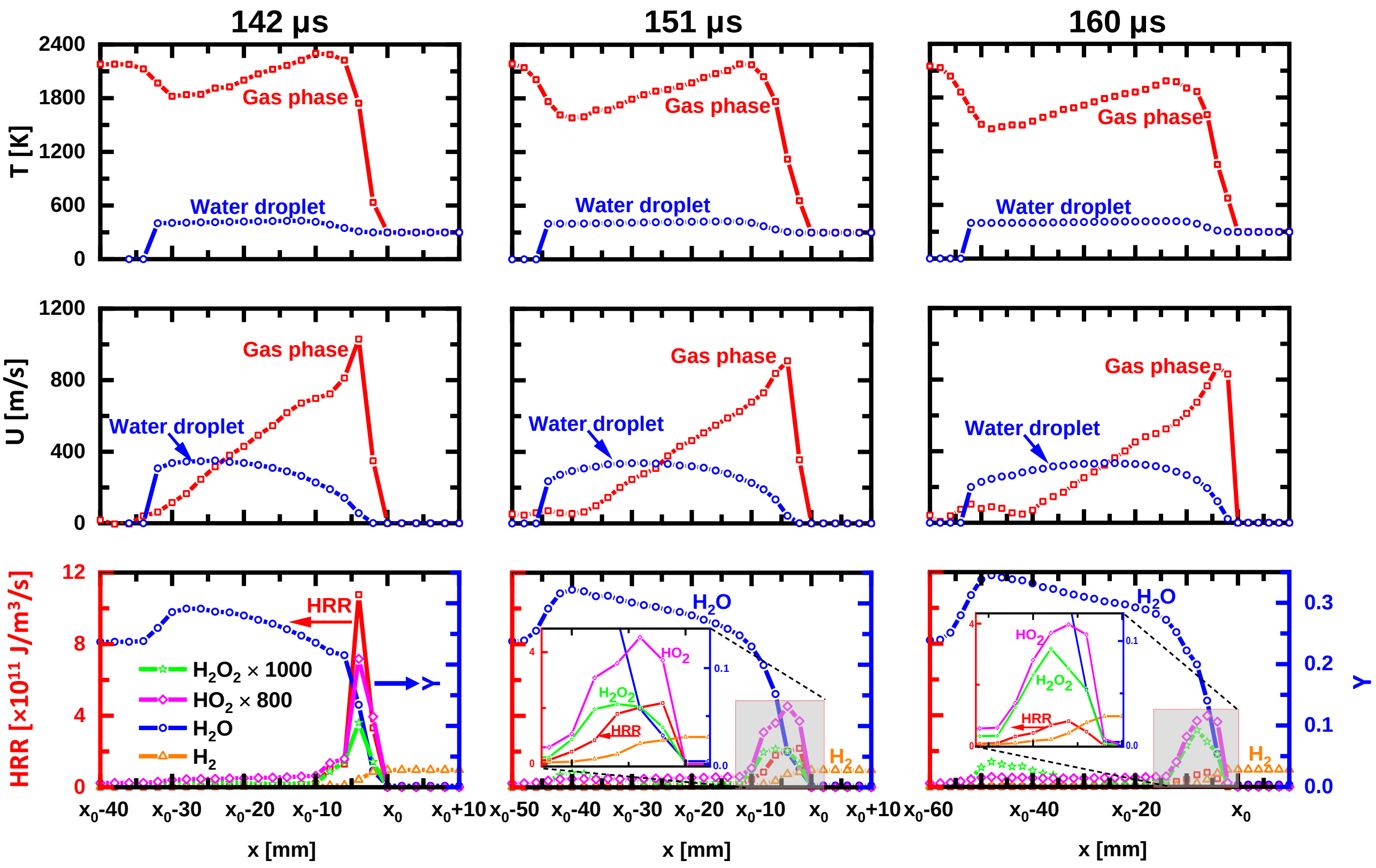

Figure 14 Spatial distributions of temperature, velocity, heat release rate, and mass fraction of main species at three instants corresponding to Fig. 12. $x_0$: the location of reaction front.

The profiles of temperatures, velocities, heat release rate, and mass fraction of main species at 142, 151 and 160 μs are depicted in Fig. 14. Note that these quantities are obtained through density-weighted averaging along the $y$-direction. When the SF and RF are gradually decoupled, the gas temperature decreases. Moreover, the droplet temperature distributions are almost not affected. The averaged gas velocity grows quickly due to the arrival of the SF and decreases to zero after about 30−40 mm after the SF. However, the peak value is gradually reduced from 142 to 160 μs. The droplet velocity slightly increases due to the momentum exchange and is close to that of the gas velocity at around $x_0$−20 mm. Based on the profiles of temperature and velocity, one can see that interphase temperature and velocity equilibria behind the DW are not reached. This is reasonable due to the spatially varying gas properties and droplet relaxation time behind the leading shock. When the detonation is quenched, the heat release decays quickly, although the length of reaction zone shows limited change (about 10 mm). The $H_2O$ mass fraction increases when the DW propagates forward although it is gradually weakened, due to more evaporating droplets behind the SF. However, $HO_2$ and $H_2O_2$ mass fractions decreases because of the weakened gas reactions. The chemical species contributions of the DW will be analysed in Section 5.2 with CEMA.

### 5.2 Chemical explosion mode in gaseous detonation

In this section, CEMA [20,21] is used to quantify the critical gas phase chemical feature in detonations. It is inspired by the Computational Singular Perturbation (CSP) method developed by Lam et al. [68–71] and has been proven a reliable tool to identify the critical combustion phenomena [20,21,72]. The differential equations for a spatially homogeneous reaction system read

$$\frac{d\boldsymbol{y}}{d\boldsymbol{t}} = \boldsymbol{\omega}(\boldsymbol{y}), \tag{30}$$

where $\boldsymbol{y}$ represents the vector of temperature and species mass fractions. $\boldsymbol{\omega}(\boldsymbol{y})$ is the chemical source term. In CEMA, eigen-analysis of the local chemical Jacobian is performed

$$\frac{d\boldsymbol{\omega}}{d\boldsymbol{t}} = \mathbf{J}_\omega \cdot \boldsymbol{\omega}(\boldsymbol{y}), \quad \mathbf{J}_\omega = \frac{\partial \boldsymbol{\omega}}{\partial \boldsymbol{y}}, \tag{31}$$

where $\mathbf{J}_{\boldsymbol{\omega}}$ is the Jacobian matrix of the chemical source term $\boldsymbol{\omega}$. The eigenmode of the Jacobian matrix associated with the eigenvalues of $\mathbf{J}_{\boldsymbol{\omega}}$, i.e., $\lambda_e = \mathbf{b}_e\ \mathbf{J}_{\boldsymbol{\omega}}\ \mathbf{a}_e$, is defined as a Chemical Explosive Mode (CEM) when the real part of $\lambda_e$ is greater than zero, i.e., $Re(\lambda_e) > 0$. It should be highlighted that $Re(\lambda_e)$ corresponds to the reciprocal timescale of the explosion $\tau_{chem}$ [20]. Here $\mathbf{a}_e$ and $\mathbf{b}_e$ are respectively the right and left eigenvectors associated with the CEM. Note that CEM is a chemical property of local gaseous mixture and indicates the propensity of ignition when the mixture is isolated (constant volume, adiabatic and droplet-free) [20]. $Re(\lambda_e) > 0$ means that the mixture can autoignite, whilst $Re(\lambda_e) < 0$ means that it is burnt or fails to auto-ignite. The condition of $Re(\lambda_e) = 0$ therefore separates the CEM region and post-combustion or inert mixing one. Moreover, the contributions of temperature or species to the CEM can be evaluated through the Explosion Index (EI) [73]

$$\mathbf{EI} = \frac{|\mathbf{a_e} \otimes \mathbf{b_e^T}|}{\sum |\mathbf{a_e} \otimes \mathbf{b_e^T}|}, \tag{32}$$

where "⊗" denotes element-wise multiplication of two vectors.

Figure 15 shows the spatial evolutions of $\lambda_{CEM} \equiv sign[Re(\lambda_e)] \cdot log_{10}[1 + |Re(\lambda_e)|]$ at 142, 151 and 160 μs in case e, corresponding to the same instants in Figs. 12 and 14. For better illustration, only CEM with positive $\lambda_{CEM}$ is shown. It is found that CEM exists between the SF and RF. This suggests that the local gaseous mixture is highly explosive. In addition, at 142 μs, the CEM behind the IW and MS are different. For the former, in the induction zone, finite value of $\lambda_{CEM}$ can only be seen immediately ahead of the RF, corresponding to the short chemical timescale $\tau_{chem}$ and hence strong reactivity. However, higher $\lambda_{CEM}$ (hence low $\tau_{chem}$) exists in the entire induction zone between the RF and MS. In this sense, the mixture behind the MS is intrinsically more explosive than behind the IW. At 151 μs, with increased localized detonation extinctions along the DW, the CEM is more distributed as more part of RF's are decoupled from the SF. At 160 μs, the RF is fully decoupled from the SF, and low values of $\lambda_{CEM}$ dominate between the RF and SF, although the high $\lambda_{CEM}$ are still seen near the RF. This implies that the chemical explosion propensity of the shocked gas is further reduced at 160 μs.

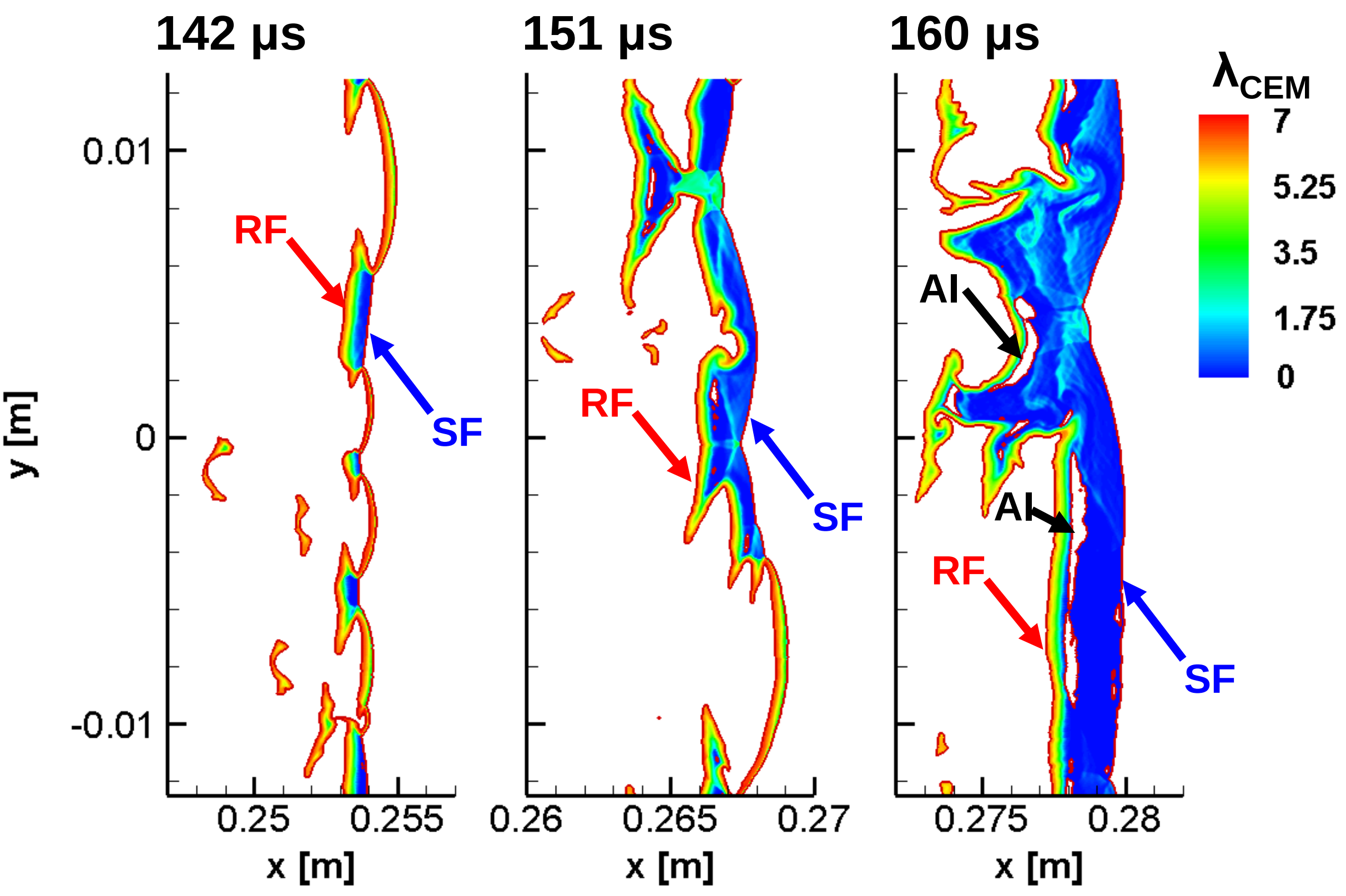

Figure 15 Distributions of the CEM in a detonation extinction process. Results from case e: $d_d^0$ = 10 μm and $z$ = 0.62. Rightmost lines: Shock Front (SF); the rest iso-lines: Reaction Front (RF); Auto-ignition: AI.

Figure 16 presents the spatial distributions of EI's for temperature and radicals which are involved non-trivially in CEM. At 142 μs, the contributions from temperature are generally important between the RF and SF. Nevertheless, at 151 and 160 μs, the EI's of radicals, such as $HO_2$ and $H_2O_2$, increase immediately behind the SF. It has been suggested that the local chemical reaction is dominated by autoignition if the radical EI is high, whilst is dominated by thermal runway if temperature plays a more important role [20]. Therefore, at 142 μs with stable detonations, thermal runaway proceeds behind the SF. Nevertheless, at 151 and 160 μs respectively with partial and global extinctions, most of the mixture in the induction zone is shown to have the propensity to auto-ignite.

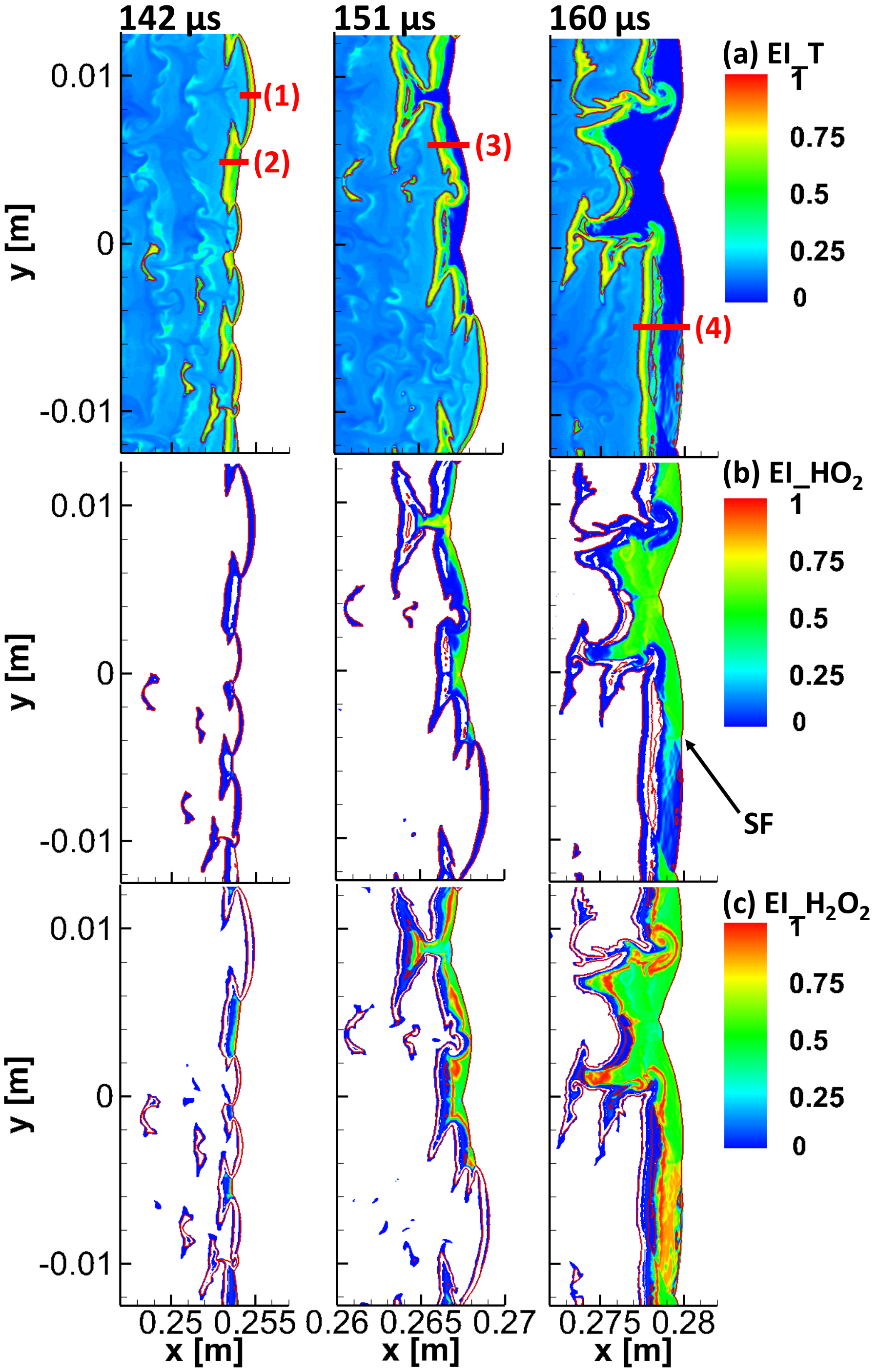

Figure 16 Distributions of EIs of (a) temperature, (b) $HO_2$, and (c) $H_2O_2$ in a detonation extinction process. Results from case e: $d_d^0$ = 10 μm and z = 0.62. Rightmost lines: shock front (SF); the rest isolines: Reaction Fronts (RF). Lines 1-4: locations of EI plot in Fig. 17.

Quantitative comparisons of temperature / species EI's in Fig. 16 are presented in Figs. 18(a)−18(d), which respectively correspond to four loci marked by the short lines 1−4 in Fig. 16(a). For each figure, the left (right) end of the *x*-axis is RF (SF) and therefore the EI variations inside the induction zone is visualized. Meanwhile, the evolutions of the chemical timescale $\tau_{chem}$ in the abovementioned four locations are demonstrated in Fig. 18. As observed in Fig. 17(a), behind the MS, the CEM is dominant by thermal runaway process in the induction zone and the chemical timescale $\tau_{chem}$ is relatively uniform (see Fig. 18). However, behind the IW (Fig. 17b), the EI's of the radical species and temperature alternate behind the SF. This is also observed in the chemical propensity within the induction zone of the pulsating *n*-heptane detonations [72]. Specifically, at $x > x_{TR1}$ = 254.08 mm and $x < x_{TR2}$ = 253.96 mm, thermal runaway occurs (temperature EI is largest). However, in between, $H_2O_2$ shows the highest contributions towards the autoignition-dominated CEM. The chemical timescale considerably decreases to 1 μs near the RF, as seen in Fig. 18. In Fig. 17(c), the region with thermal runaway CEM is considerably reduced and autoignition CEM is present at $x > x_{TR3}$ = 266.64 mm. For most of the induction zone ($x > x_{TR3}$), $\tau_{chem}$ varies between 0.01 and 10 μs. In Fig. 17(d) in which complete RF/SF decoupling occurs, the thermal runaway region ($x < x_{TR5}$ = 278.54 mm) is further reduced.

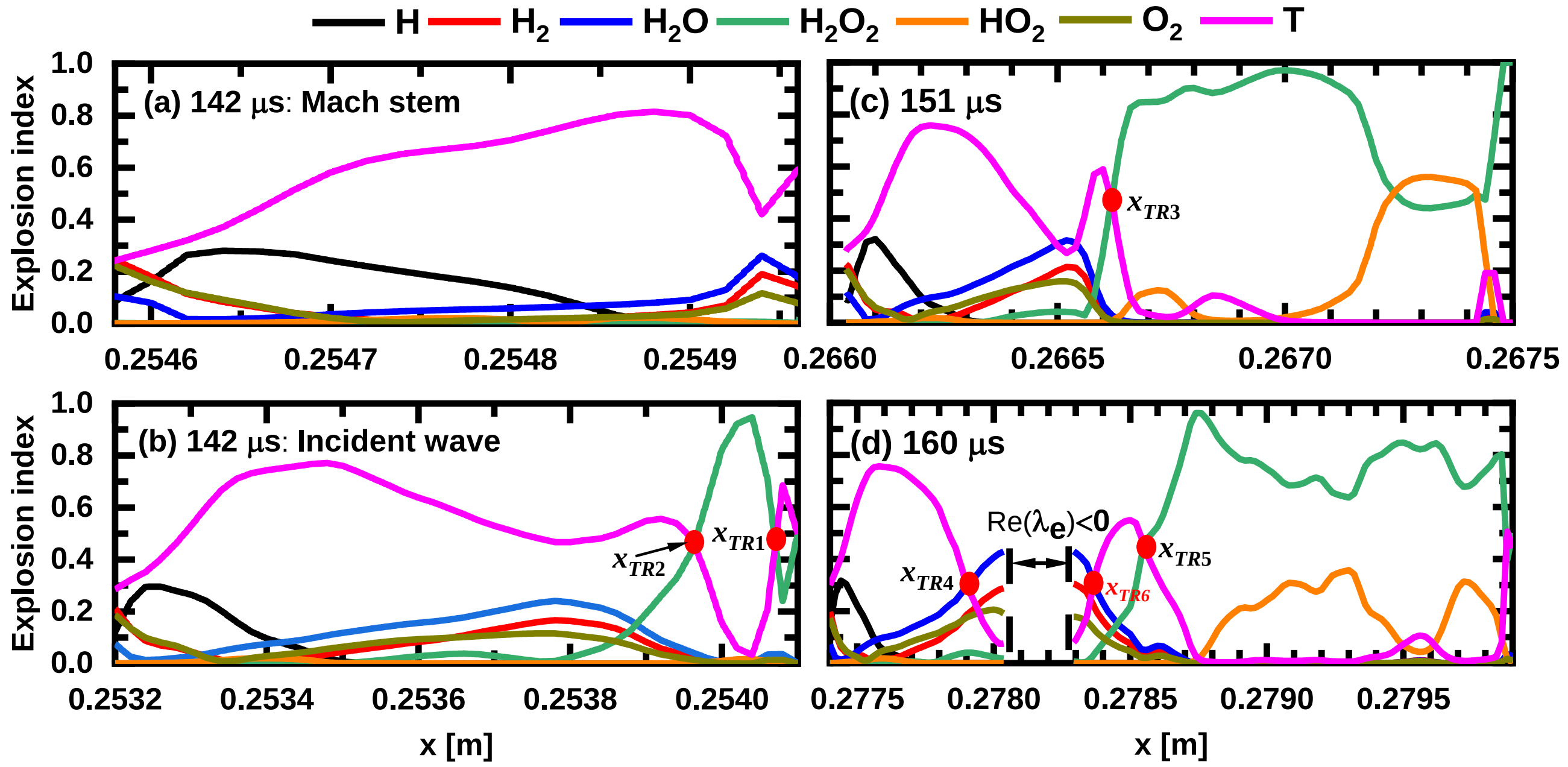

Figure 17 Spatial distributions of EI's in a detonation extinction process along the lines 1−4 in Fig. 16. Red dots: transition loci for thermal runaway and radical explosion.

Interestingly, the region with Re($\lambda_e$) < 0 in Figs. 17(d) and 18 is the localized burned island near the RF and surrounded by the mixture with autoignition CEM, as marked with "AI" in Fig. 15. This is caused by the continuously increased induction zone length between the RF and SF in detonation extinction process, which indicates a longer residence time of the explosive mixture in the shocked area. Besides, radical back diffusion from the RF may also promote the onset of localized AI spots. However, probably due to considerably reduced gas temperature (< 1,200 K, see Fig. 12) in the induction zone, further development of these AI spots into reaction front propagation towards the leading SF is not observed. This is different from the evolutions of the local explosions ahead of the travelling RF in one-dimensional *n*-heptane/air detonations, which induces periodic coupling of the SF and RF and hence pulsating detonations [72]. Beyond 160 μs, the AI spots are further extended along the spanwise direction before the RF, which may further weaken the RF propagation due to partial reaction of the mixture before the SF.

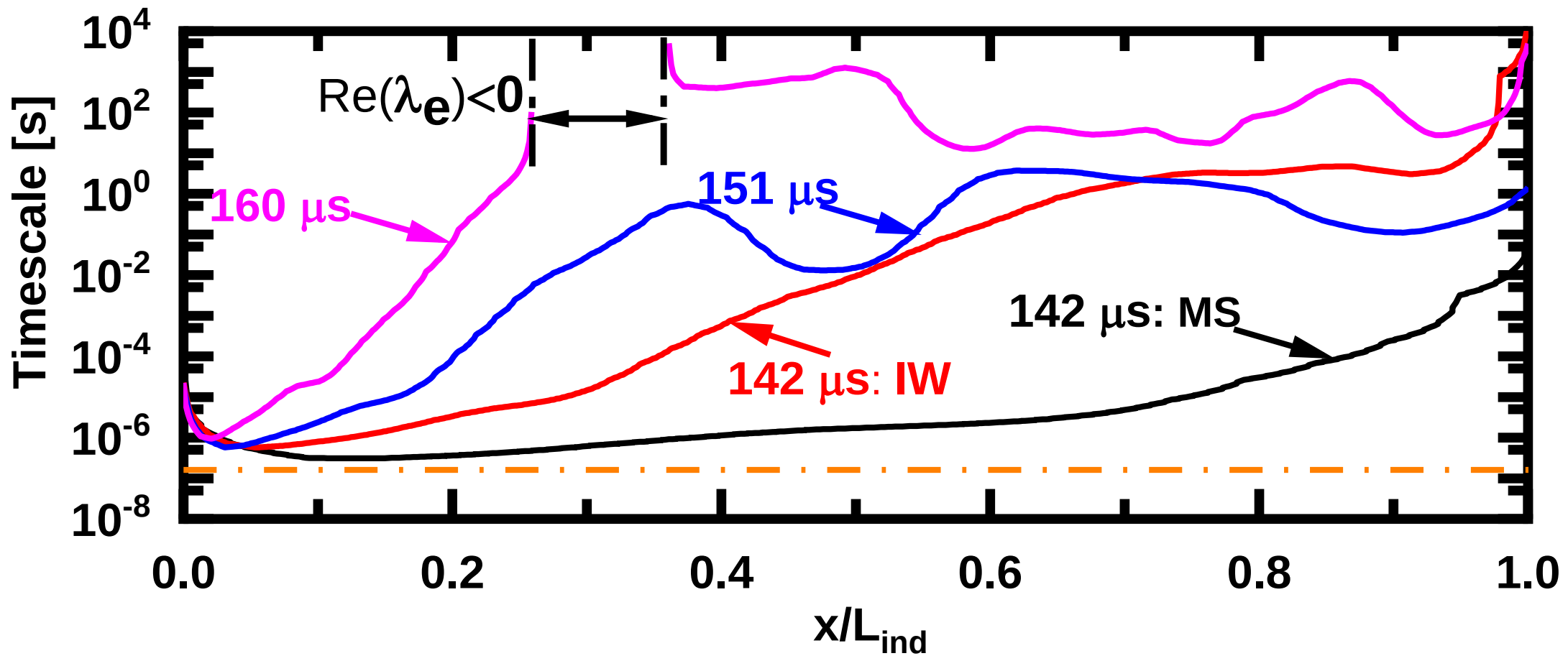

Figure 18 Spatial distributions of chemical timescale within the induction zone. Dashed line: timescale estimated from induction zone length divided by Chapman–Jouguet speed. $x$ = 0: reaction front; $x$ = 1.0: shock front.

### 5.3 Interaction between detonation and water sprays

To clarify the effects of fine water sprays on hydrogen detonation, interphase exchange of mass, momentum, and energy will be discussed in this section. The density-weighted average interphase transfer rates ($S_{mass}$, $S_{mom}$ and $S_{energy}$ in Eqs. 24−26) are presented in Fig. 19. A positive mass (energy and momentum) transfer rate indicates that the transfer from liquid (gas) phase to gas (liquid) phase. To explore the water mass loading effects, four cases with stable detonation propagation ($z$ = 0.031 – 0.465) are first discussed. The results in Fig. 19 correspond to an instant when the DW's lie at the end of the two-phase section ($x_0$ = 0.28 m). In general, the transfer rates increase with the water droplet mass loading. Note that although the magnitudes of the momentum exchange are rapidly reduced after the SF, kinetic equilibrium is not reached in the detonated gas and at about $x_0 - 20$ mm momentum transfer rate becomes negative, indicating that the momentum transfers from the droplet phase to gas phase. This can be seen clearly in the inset of Fig. 19(c) and is also unveiled from the velocity profiles in Fig. 14.

It can be observed that energy and momentum exchanges start immediately at $x_0$ (i.e., SF), but pronounced mass transfer (i.e., droplet evaporation) occurs at $x_0 - 5$ mm, well behind both SF and RF. Consequently, the energy and/or momentum transfer (convective heat transfer and/or drag force) are expected to have more direct influence on the RF and SF than the mass transfer (water vapour addition and evaporative cooling). Specifically, attenuation of the leading shock would occur for accelerating the dispersed water droplets. However, it would take a longer distance to have pronounced shock attenuation by gas−droplet interactions [74]. For instance, in case b2 ($z$ = 0.31) discussed above, the intensity of the leading SF is $15.4p_0$ when it exits from the domain, only 18.2% lower than that when the DW enters the two-phase section. Furthermore, since there is limited droplet evaporation near RF, the chemical reaction behind the shock is considerably affected by the convective heat transfer between the gas and water sprays. Nevertheless, the averaged energy transfer rate further increases when the droplet evaporation becomes strong at around $x_0 - 10$ mm, due to the absorption of the latent heat. Therefore, although mass transfer occurs after the SF and RF, however, it can still indirectly weaken the gas temperature around the RF through heat conduction. This can also be observed in Figs. 14.

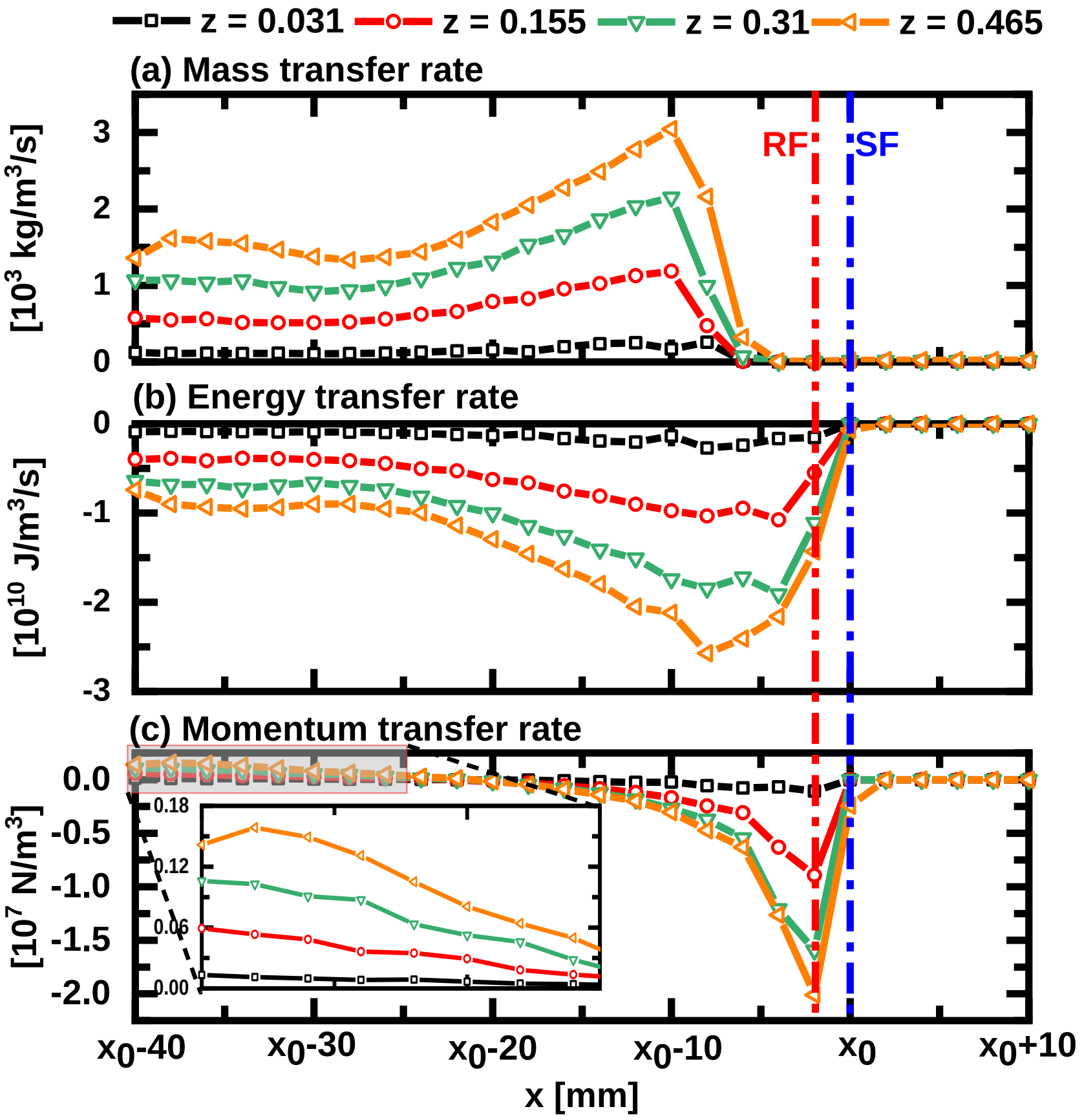

Figure 19 Profiles of averaged transfer rates of (a) mass, (b) energy and (c) momentum with various mass loadings. $d_d^0$ = 10 μm. $x_0$: leading shock front. RF: reaction front.

The volume-averaged interphase transfer rates in stable detonations subject to different droplet mass loadings and droplet diameters are demonstrated in Fig. 20. In all the cases, volume averaging is performed from one instant when the DW's are near the exit of the two-phase section. One can see from Fig. 20 that for the same droplet size (e.g., 10 μm), the rates of mass, energy, and momentum increase with the water mass loading. Moreover, under the same mass loading (e.g., 0.1), the smaller the droplet size, the larger the interphase exchange rates. It is also seen from Fig. 20(c) that the droplet size is shown to have a limited influence on the interphase momentum exchange, probably due to relatively small droplet diameters.

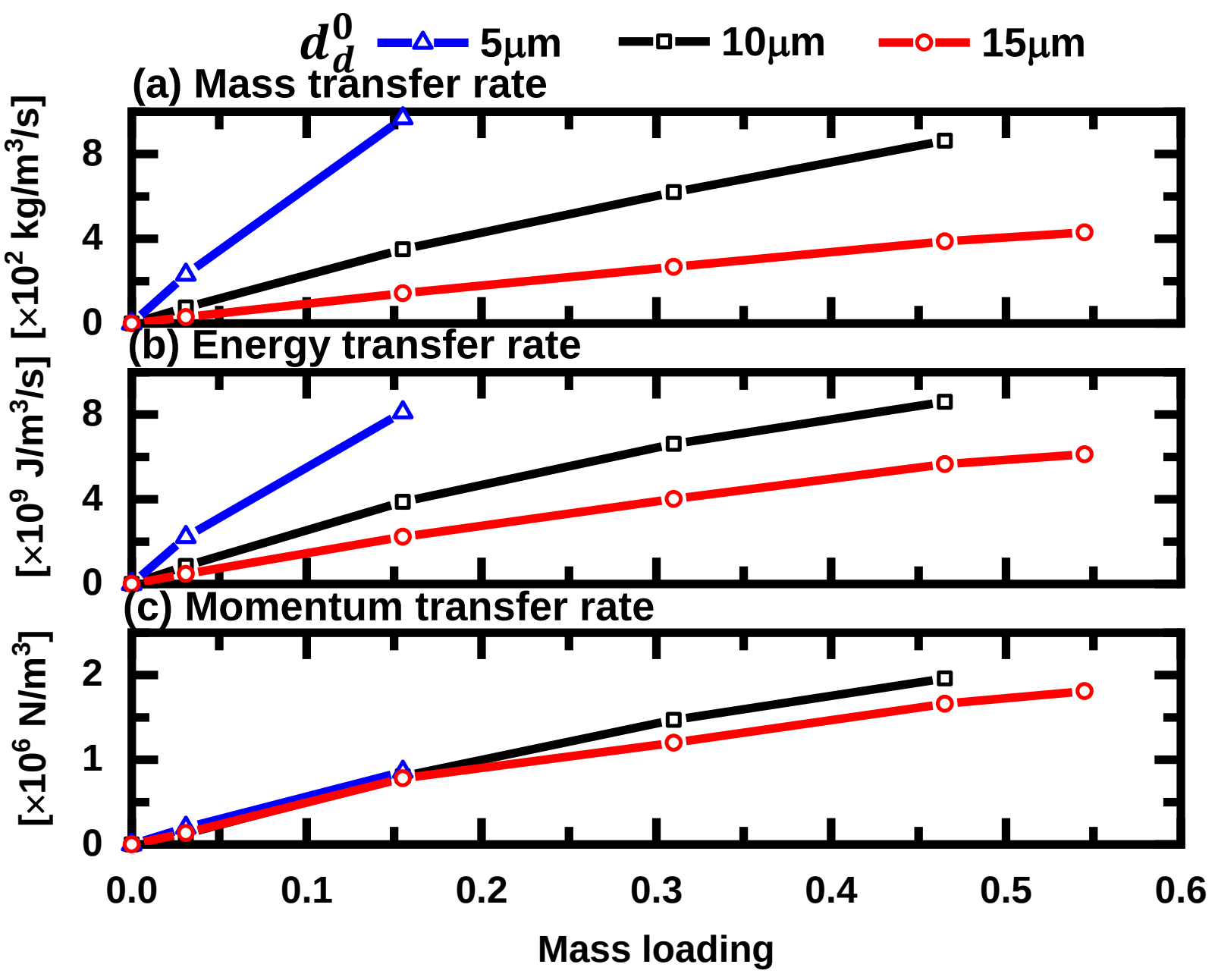

Figure 20 Volume-averaged transfer rates of (a) mass, (b) energy, and (c) momentum as functions of droplet mass loading and diameter.

The evolutions of volume-averaged interphase transfer rates and HRR in detonation extinction process (case e) are shown in Fig. 21. For comparison, the counterpart results from a stable detonation case (case b2) are also presented. One can see from Fig. 21(a) that the mass and energy transfer rates monotonically increase when the DW is transmitted into the two-phase section. This is because more water droplets enter the detonated area. The energy transfer rate levels off between 130 μs and 150 μs, between which the HRR starts to decrease, signifying the RF/SF decoupling at some DW locations. It is seen that the energy exchange is high for the entire extinction process and gradually decreases at 150 μs, which is close to the ultimate SF/RF decoupling point (153 μs). The mass transfer rate starts to decrease at 160 μs. These tendencies are different from the results of the stable detonation in Fig. 21(b), in which the relatively constant, albeit fluctuating, HRR and monotonic increases of interphase exchange rates are observable.

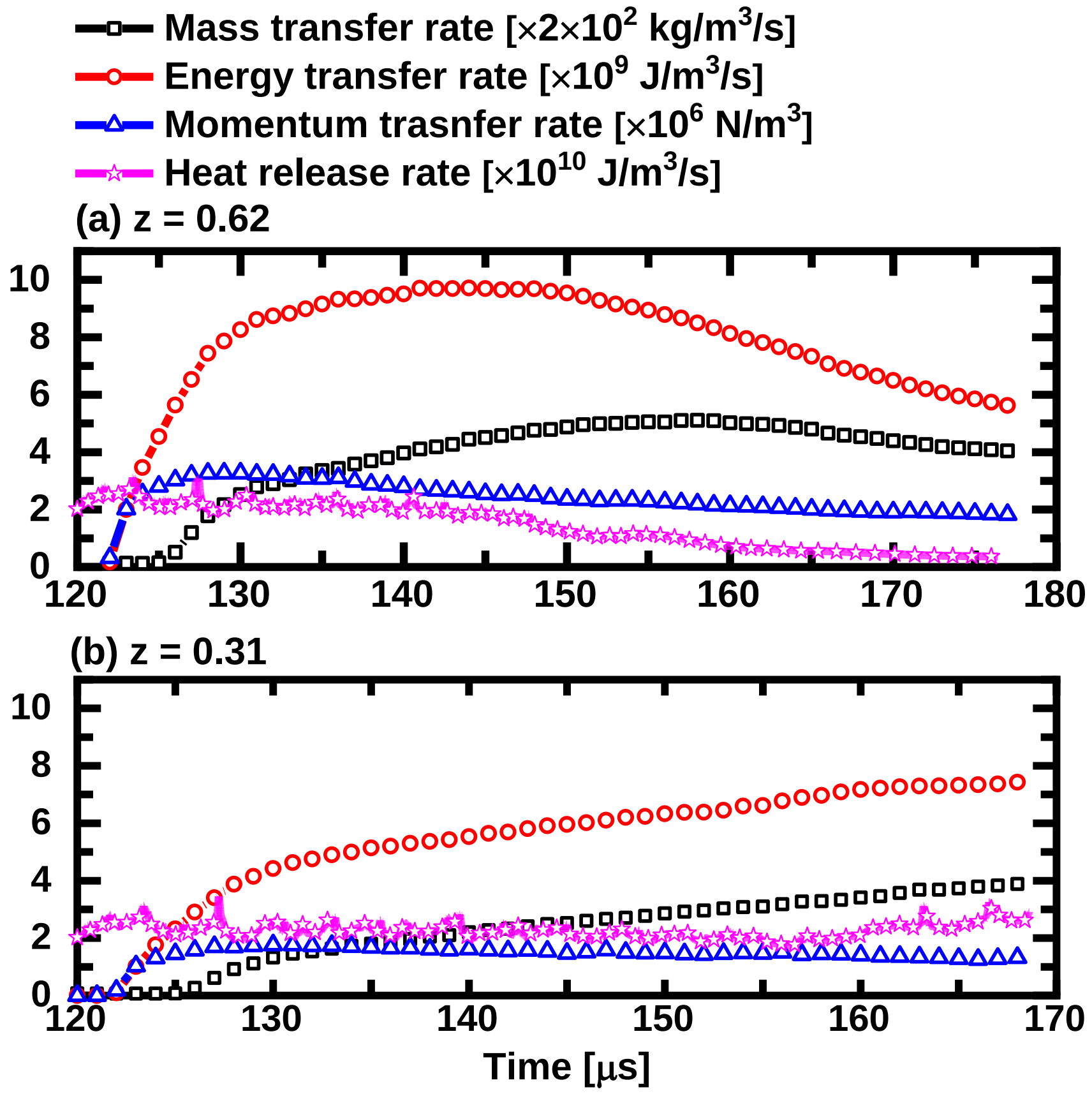

Figure 21 Time history of volume-averaged interphase transfer rates and combustion HRR with mass loading of (a) 0.62 and (b) 0.31. $d_d^0$ = 10 μm.

Figure 22 further shows the coupling between the DW's and local water sprays in cases e and b2. They are averaged from the domain with full width and 5 mm thickness centring at the RF, thereby covering the RF−SF complex. For both cases, the mass transfer rate is almost zero because the droplet temperatures are still low and this domain lies before the EOF. The energy and momentum transfer rates are almost constant when the detonation is stable, indicating the overall balance between the RF−SF complex and water droplets. However, monotonic decrease of the two rates starts at around 123 μs, much earlier than those from the counterpart results when the entire domain is considered.

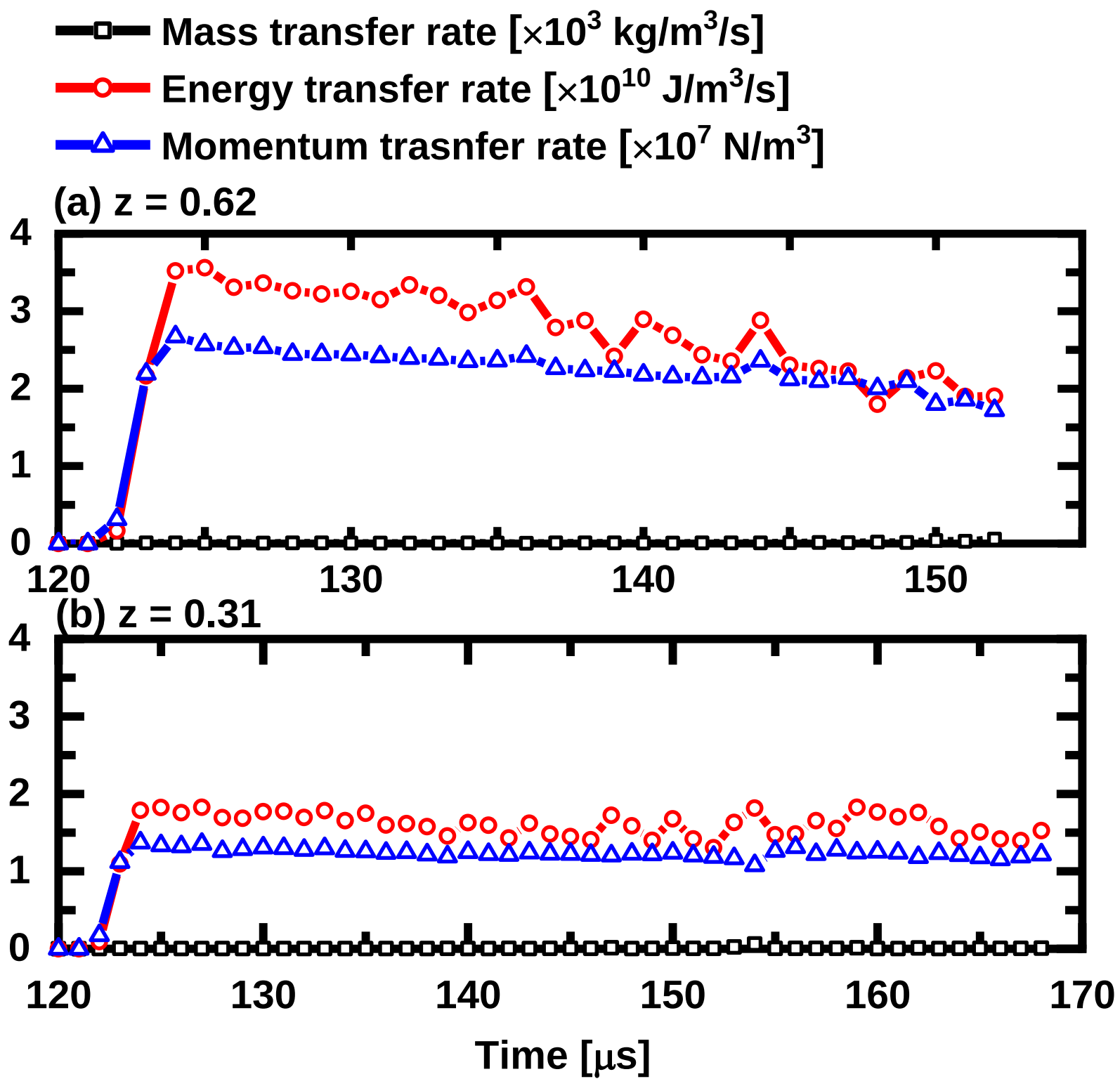

Figure 22 Time history of volume-averaged interphase transfer rates near the reaction front with mass loading of (a) 0.62 and (b) 0.31. $d_d^0$ = 10 μm.

Table 2 Numerical experiments for interactions between detonation wave and water droplets

| Case | Drag force | Convective heat transfer | Droplet evaporation and latent heat absorption |
|---|---|---|---|
| e | √ | √ | √ |
| e1 | √ | √ | × |
| e2 | √ | × | × |

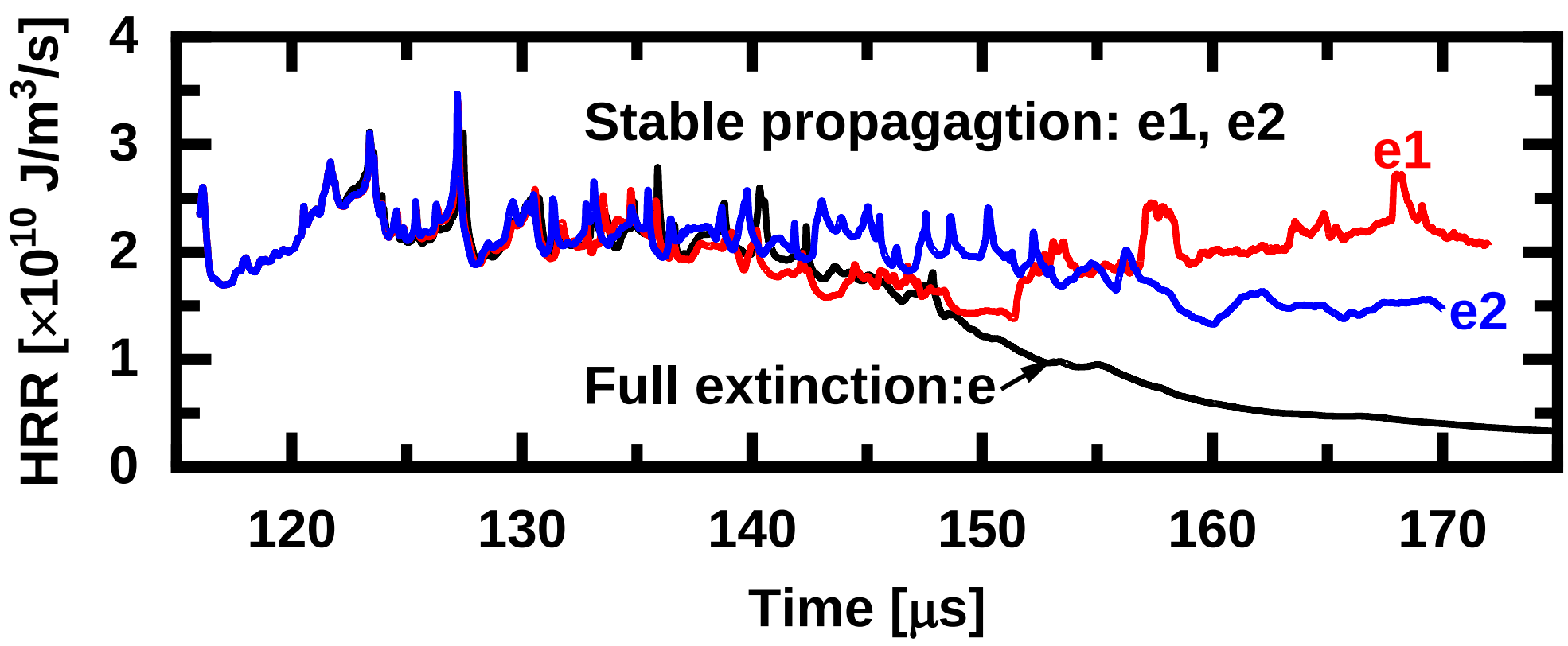

Figure 23 Time history of volume-averaged heat release rate: effects of mass, momentum, and energy transfer (z = 0.62 and $d_d^0$ = 10 μm).

To further clarify the effects of water droplets on detonation extinction, numerical experiments are performed through switching off the respective models for the two-phase interactions. They are based on case e. As tabulated in Table 2, e1 consider interphase momentum and energy exchanges (excluding droplet evaporation and latent heat absorption), whilst e2 only includes momentum exchange. The time histories of averaged HRR from these cases are illustrated in Fig. 23. It is observed that interphase coupling appreciably affects the DW propagation and their influences become pronounced at 135 μs (about 13 μs after the DW is transmitted in the water sprays). As discussed in Section 5.1, the DW in case e is fully extinguished. When the evaporation model is disabled (therefore mass transfer and latent heat absorption is not considered), the DW in case e1 can stably propagate in the whole two-phase section. This can be also confirmed through the corresponding contours of gas temperature and HRR (not shown here). This implies that the droplet evaporation and accompanied heat absorption are important for inducing DW extinction. However, due to the finite distance between the evaporating zone and DW as indicated in Fig. 19, the effects are indirect, but may be intensified when the DW is gradually weakened due to the reduced distinctions between diffusion and chemical (see Fig. 18) timescales.

Moreover, if both droplet evaporation and convective heat transfer are deactivated, stable DW is also observed in case e2. This corroborates the roles of interphase heat and mass exchanges in detonation inhibition, through comparisons of e2 and e. Interestingly, after 155 μs, HRR in case e2 is lower than that in e1. Based on our results, the momentum exchange between gas and (non-evaporating / non-heated) droplets in e2 is generally higher than that in e1, which may lead to stronger attenuation effects on the leading SF. Therefore, the weaker leading SF in e2 can be expected, which accordingly leads to weaker chemical reactions behind it. Overall, the roles of the interphase coupling of energy and mass in hydrogen detonation inhibition can be further confirmed through the foregoing numerical experiments. This is different from the observations in Refs. [12,13], in which the momentum extraction from the gas phase is highlighted. This may be because relatively large water droplets are considered in their work, as indicated in Fig. 1.

## 6. Conclusions

Extinction of two-dimensional hydrogen/air detonations in fine water sprays is computationally studied with a hybrid Eulerian-Lagrangian method considering two-way gas−liquid coupling. A detonation extinction map parameterized by water droplet size and mass loading is achieved and shows whether the gaseous detonation can stably propagate in water mists depends on the water mass loading and initial droplet size. General features of gas phase and liquid droplets and detailed detonation structures are well captured. The influence of water droplet mass loading on the hydrogen detonation is examined by the trajectories of peak pressure, and the numerical results indicate that the increased mass loading leads to detonation extinction. Meanwhile, the larger the mass loading, the earlier the extinction occurs in the two-phase section. Furthermore, smaller water droplet size would also lead to extinction of the detonation waves.

Detonation extinction analysis is performed with the evolutions of detonation frontal structure, the spatial distribution of thermochemical variables and interphase transfer rates. It is shown that the droplets need longer time to be heated and the evaporation becomes weaker behind the EOF due to the spatial delay of the RF relative to the SF. No chemical reactions proceed at these triple points and meanwhile their pressure superposition does not induce chemical reactions behind MS. The distributions of temperature and velocity indicate that interphase temperature and velocity equilibria behind the DW are not reached due to the spatially varying gas properties and droplet relaxation time behind the leading shock. In detonation extinction process, heat release and $HO_2$/$H_2O_2$ mass fractions are reduced, while $H_2O$ mass fraction increases. Moreover, the detailed information of the chemical reaction in the detonation extinction process is analyzed with the chemical explosive mode analysis. The analysis confirms that the shocked gas in the induction zone is highly explosive. For stable detonations, thermal runaway dominates CEM behind the Mach stem, whereas for those behind the incident wave, auto-ignition and temperature play alternately dominant roles. When the induction zone increases as the reaction front and shock front are decoupled, localized burned pockets surrounded by the autoignition CEM can be observed.

In addition, the interactions between the detonation wave and water droplets are discussed. The energy and momentum transfer would have stronger influence on the shock front and reaction front than the mass transfer, which starts well behind the detonation wave. It is also found that the interphase exchange rates increase with the water mass loading. Under the same mass loading, the smaller the droplet size, the larger the interphase transfer rates. The size of the fine water droplets has a limited influence on the interphase momentum exchange. Moreover, high energy and mass transfer rates are observed at the onset of detonation extinction, and they gradually decrease when the reaction and detonation fronts are decoupled.

**Acknowledgements**

This work used the computational resources of ASPIRE 1 Cluster in The National Supercomputing Centre, Singapore (https://www.nscc.sg). Yong Xu is supported by the NUS Research Scholarship. Professor Zhuyin Ren and Dr Wantong Wu at Tsinghua University are thanked for sharing the CEMA subroutines.

## References

[1] E.S. Oran, G. Chamberlain, A. Pekalski, Mechanisms and occurrence of detonations in vapor cloud explosions, Prog. Energy Combust. Sci. 77 (2020).

[2] G. Atkinson, E. Cowpe, J. Halliday, D. Painter, A review of very large vapour cloud explosions: Cloud formation and explosion severity, J. Loss Prev. Process Ind. 48 (2017) 367–375.

[3] C. Catlin, Passive explosion suppression by blast-induced atomisation from water containers, J. Hazard. Mater. 94 (2002) 103–132.

[4] G.O. Thomas, On the conditions required for explosion mitigation by water sprays, Process Saf. Environ. Prot. 78 (2000) 339–354.

[5] G. Grant, J. Brenton, D. Drysdale, Fire suppression by water sprays, Prog. Energy Combust. Sci. 26 (2000) 79–130.

[6] R. Zheng, K. Bray, B. Rogg, Effect of Sprays of Water and NaCl-Water Solution on the Extinction of Laminar Premixed Methane-Air Counterflow Flames, Combust. Sci. Technol. 126 (1997) 389–401.

[7] Y. Song, Q. Zhang, Quantitative research on gas explosion inhibition by water mist, J. Hazard. Mater. 363 (2019) 16–25.

[8] L.R. Boeck, A. Kink, D. Oezdin, J. Hasslberger, T. Sattelmayer, Influence of water mist on flame acceleration, DDT and detonation in H2-air mixtures, Int. J. Hydrogen Energy. 40 (2015) 6995–7004.

[9] G. Jourdan, L. Biamino, C. Mariani, C. Blanchot, E. Daniel, J. Massoni, L. Houas, R. Tosello, D. Praguine, Attenuation of a shock wave passing through a cloud of water droplets, Shock Waves. 20 (2010) 285–296.

[10] A. Chauvin, G. Jourdan, E. Daniel, L. Houas, R. Tosello, Experimental investigation of the propagation of a planar shock wave through a two-phase gas-liquid medium, Phys. Fluids. 23 (2011) 1–13.

[11] K.C. Adiga, H.D. Willauer, R. Ananth, F.W. Williams, Implications of droplet breakup and formation of ultra fine mist in blast mitigation, Fire Saf. J. 44 (2009) 363–369.

[12] R. Ananth, H.D. Willauer, J.P. Farley, F.W. Williams, Effects of Fine Water Mist on a Confined Blast, Fire Technol. 48 (2012) 641–675.

[13] D.A. Schwer, K. Kailasanath, Numerical simulations of the mitigation of unconfined explosions using water-mist, Proc. Combust. Inst. 31 II (2007) 2361–2369.

[14] G.O. Thomas, M.J. Edwards, D.H. Edwards, Studies of detonation quenching by water sprays, Combust. Sci. Technol. 71 (1990) 233–245.

[15] U. Niedzielska, L.J. Kapusta, B. Savard, A. Teodorczyk, Influence of water droplets on propagating detonations, J. Loss Prev. Process Ind. 50 (2017) 229–236.

[16] G. Jarsalé, F. Virot, A. Chinnayya, Ethylene–air detonation in water spray, Shock Waves. 26 (2016) 561–572.

[17] H. Watanabe, A. Matsuo, K. Matsuoka, A. Kawasaki, J. Kasahara, Numerical investigation on propagation behavior of gaseous detonation in water spray, Proc. Combust. Inst. 37 (2019) 3617–3626.

[18] H. Watanabe, A. Matsuo, A. Chinnayya, K. Matsuoka, A. Kawasaki, J. Kasahara, Numerical analysis of the mean structure of gaseous detonation with dilute water spray, J. Fluid Mech. 887 (2020).

[19] H. Watanabe, A. Matsuo, A. Chinnayya, K. Matsuoka, A. Kawasaki, J. Kasahara, Numerical analysis on behavior of dilute water droplets in detonation, Proc. Combust. Inst. 000 (2020) 1–8.

[20] T.F. Lu, C.S. Yoo, J.H. Chen, C.K. Law, Three-dimensional direct numericaal simulation of a turbulent lifted hydrogen jet flame in heated coflow: A chemical explosive mode analysis, J. Fluid Mech. 652 (2010) 45–64.

[21] W. Wu, Y. Piao, Q. Xie, Z. Ren, Flame diagnostics with a conservative representation of

chemical explosive mode analysis, AIAA J. 57 (2019) 1355–1363.
[22] W. Sutherland, LII. The viscosity of gases and molecular force , London, Edinburgh, Dublin Philos. Mag. J. Sci. 36 (1893) 507–531.
[23] B.E. Poling, J.M. Prausnitz, J.P. O'connell, The properties of gases and liquids, Mcgraw-hill New York, 2001.
[24] B. Mcbride, S. Gordon, M. Reno, Coefficients for Calculating Thermodynamic and Transport Properties of Individual Species, National Aeronautics and Space Administration, 1993.
[25] G.B. Macpherson, N. Nordin, H.G. Weller, Particle tracking in unstructured, arbitrary polyhedral meshes for use in CFD and molecular dynamics, Commun. Numer. Methods Eng. 25 (2009) 263–273.
[26] C.T. Crowe, J.D. Schwarzkopf, M. Sommerfeld, Y. Tsuji, Multiphase flows with droplets and particles, CRC Press, New York, U.S., 1998.
[27] R.H. Perry, D.W. Green, J.O. Maloney, Perry's chemical engineers' handbook, 7th Editio, 1998.
[28] Bradley RS, F. NABT-E, D. in GM, Evaporation and droplet growth in gaseous media, 1959.
[29] S.S. Sazhin, Advanced models of fuel droplet heating and evaporation, Prog. Energy Combust. Sci. 32 (2006) 162–214.
[30] Z. Huang, M. Zhao, H. Zhang, Modelling n-heptane dilute spray flames in a model supersonic combustor fueled by hydrogen, Fuel. 264 (2020) 116809.
[31] W.E. Ranz, W.R. Marshall, Evaporation from Drops, Part I., Chem. Eng. Prog. 48 (1952) 141–146.
[32] E.N. Fuller, P.D. Schettler, J.C. Giddings, A New Method for Prediction of Binary Gas - Phase Diffusion Coefficients, Ind. Eng. Chem. 58 (1966) 18–27.
[33] E.L. Cussler, Diffusion: Mass Transfer in Fluid Systems, 3rd Edition, 2009.
[34] C. Crowe, J. Schwarzkopf, M. Sommerfeld, Y. Tsuji, Multiphase Flows with Droplets and Particles, 2nd ed., CRC Press, 2011.
[35] A.B. Liu, D. Mather, R.D. Reitz, Modeling the effects of drop drag and breakup on fuel sprays, SAE Tech. Pap. 1 (1993).
[36] S. Cheatham, K. Kailasanath, Numerical modelling of liquid-fuelled detonations in tubes, Combust. Theory Model. 9 (2005) 23–48.
[37] C.B. Henderson, Drag coefficients of spheres in continuum and rarefied flows, AIAA J. 14 (1976) 707–708.
[38] O. Igra, K. Takayama, Shock tube study of the drag coefficient of a sphere in a nonstationary flow, Shock Waves. (1992) 491–497.
[39] G. Tedeschi, H. Gouin, M. Elena, Motion of tracer particles in supersonic flows, Exp. Fluids. 26 (1999) 288–296.
[40] Z. Huang, M. Zhao, Y. Xu, G. Li, H. Zhang, Eulerian-Lagrangian modelling of detonative combustion in two-phase gas-droplet mixtures with OpenFOAM: Validations and verifications, Fuel. 286 (2021) 119402.
[41] H.G. Weller, G. Tabor, H. Jasak, C. Fureby, A tensorial approach to computational continuum mechanics using object-oriented techniques, Comput. Phys. 12 (1998) 620.
[42] C.J. Greenshields, H.G. Weller, L. Gasparini, J.M. Reese, Implementation of semi-discrete, non-staggered central schemes in a colocated, polyhedral, finite volume framework, for high-speed viscous flows, Int. J. Numer. Methods Fluids. 63 (2010) 1–21.
[43] H. Zhang, M. Zhao, Z. Huang, Large eddy simulation of turbulent supersonic hydrogen flames with OpenFOAM, Fuel. Submitted (2020) 118812.
[44] Z. Huang, M. Zhao, Y. Xu, G. Li, H. Zhang, Eulerian-Lagrangian modelling of detonative combustion in two-phase gasdroplet mixtures with OpenFOAM, Fuel. Under revi (2020).
[45] M. Zhao, J.M. Li, C.J. Teo, B.C. Khoo, H. Zhang, Effects of Variable Total Pressures on Instability and Extinction of Rotating Detonation Combustion, Flow, Turbul. Combust. 104 (2020) 261–290.
[46] M. Zhao, H. Zhang, Origin and chaotic propagation of multiple rotating detonation waves in

hydrogen/air mixtures, Fuel. 275 (2020) 117986.
[47] M. Zhao, H. Zhang, Modelling rotating detonative combustion fueled by partially pre-vaporized n-heptane sprays, Submitted to Fuel, (2021).
[48] M. Zhao, H. Zhang, Large eddy simulation of non-reacting flow and mixing fields in a rotating detonation engine, Fuel. 280 (2020) 118534.
[49] Q. Meng, M. Zhao, H. Zheng, H. Zhang, Eulerian − Lagrangian modelling of rotating detonative combustion in partially pre-vaporized n -heptane sprays with hydrogen addition, Fuel. 119808 (2020).
[50] A. Kurganov, S. Noelle, G. Petrova., Semidiscrete central-upwind schemes for hyperbolic conservation laws and Hamilton - Jacobi equations, SIAM J. Sci. Comput. 23 (2001) 707–740.
[51] M. Ó Conaire, H.J. Curran, J.M. Simmie, W.J. Pitz, C.K. Westbrook, A comprehensive modeling study of hydrogen oxidation, Int. J. Chem. Kinet. 36 (2004) 603–622.
[52] Y. Mahmoudi, K. Mazaheri, High resolution numerical simulation of triple point collision and origin of unburned gas pockets in turbulent detonations, Acta Astronaut. 115 (2015) 40–51.
[53] K. Mazaheri, Y. Mahmoudi, M.I. Radulescu, Diffusion and hydrodynamic instabilities in gaseous detonations, Combust. Flame. 159 (2012) 2138–2154.
[54] S. Yungster, K. Radhakrishnan, Pulsating one-dimensional detonations in hydrogen-air mixtures, Combust. Theory Model. 8 (2004) 745–770.
[55] Z.. Liu, A.K. Kim, A Review of water mist fire suppression systems – fundamental studies, J. Fire Prot. Eng. 10 (2000) 32–50.
[56] R.W. Houim, R.T. Fievisohn, The influence of acoustic impedance on gaseous layered detonations bounded by an inert gas, Combust. Flame. 179 (2017) 185–198.
[57] H. Watanabe, A. Matsuo, K. Matsuoka, A. Kawasaki, J. Kasahara, Numerical investigation on propagation behavior of gaseous detonation in water spray, Proc. Combust. Inst. 37 (2019) 3617–3626.
[58] Y. Mahmoudi, K. Mazaheri, S. Parvar, Hydrodynamic instabilities and transverse waves in propagation mechanism of gaseous detonations, Acta Astronaut. 91 (2013) 263–282.
[59] M.H. Lefebvre, E.S. Oran, Analysis of the shock structures in a regular detonation, Shock Waves. 4 (1995) 277–283.
[60] S.O. Saunders, Experimental observations of the transition to detonation in an explosive gas, Proc. R. Soc. London. Ser. A. Math. Phys. Sci. 295 (1966) 13–28.
[61] A.I. Gavrikov, A.A. Efimenko, S.B. Dorofeev, A model for detonation cell size prediction from chemical kinetics, Combust. Flame. 120 (2000) 19–33.
[62] R.A. Strehlow, R.E. Maurer, S. Rajan, Transverse waves in detonations. I: Spacing in the hydrogen-oxygensystem, AIAA J. 7 (1969) 323–328.
[63] R.K. Kumar, Detonation cell widths in hydrogenoxygendiluent mixtures, Combust. Flame. 80 (1990) 157–169.
[64] V.N. Gamezo, D. Desbordes, E.S. Oran, Formation and evolution of two-dimensional cellular detonations, Combust. Flame. 116 (1999) 154–165.
[65] E.S. Oran, J.W. Weber, E.I. Stefaniw, M.H. Lefebvre, J.D. Anderson, A numerical study of a two-dimensional H2-O2-Ar detonation using a detailed chemical reaction model, Combust. Flame. 113 (1998) 147–163.
[66] K. Mazaheri, Y. Mahmoudi, M. Sabzpooshani, M.I. Radulescu, Experimental and numerical investigation of propagation mechanism of gaseous detonations in channels with porous walls, Combust. Flame. 162 (2015) 2638–2659.
[67] S. Taileb, J. Melguizo-Gavilanes, A. Chinnayya, Influence of the chemical modeling on the quenching limits of gaseous detonation waves confined by an inert layer, Combust. Flame. 218 (2020) 247–259.
[68] S.H. Lam, Singular Perturbation for Stiff Equations Using Numerical Methods., in: B.C. In: Casci C. (Ed.), Recent Adv. Aerosp. Sci., Springer, Boston, MA., 1985: pp. 3–19.
[69] S.H. Lam, D.A. Goussis, The CSP method for simplifying kinetics, Int. J. Chem. Kinet. 26

(1994) 461–486.
[70] S.H. Lam, Using CSP to Understand Complex Chemical Kinetics, Combust. Sci. Technol. 89 (1993) 375–404.
[71] S.H. Lam, Reduced chemistry-diffusion coupling, Combust. Sci. Technol. 179 (2007) 767–786.
[72] M. Zhao, Z. Ren, H. Zhang, Pulsating detonative combustion in n-heptane/air mixtures under off-stoichiometric conditions, Combust. Flame. 226 (2021) 285–301.
[73] Z. Luo, C.S. Yoo, E.S. Richardson, J.H. Chen, C.K. Law, T. Lu, Chemical explosive mode analysis for a turbulent lifted ethylene jet flame in highly-heated coflow, Combust. Flame. 159 (2012) 265–274.
[74] Z. Huang, H. Zhang, On the interactions between a propagating shock wave and evaporating water droplets, Phys. Fluids. 32 (2020) 123315.